\documentclass{ar-1col-S2O}
\usepackage[numbers]{natbib}
\usepackage{url}
\usepackage{amsmath}
\usepackage{braket}
\jname{Template for Annual Review of Nuclear and Particle Science}
\jvol{Vol. XX}
\jyear{2025}

\begin{document}

\markboth{Dean Lee}{Lattice EFT}

\title{Lattice Effective Field Theory Simulations of Nuclei}

\author{Dean Lee,$^1$
\affil{$^1$Facility for Rare Isotope Beams and Department of Physics and Astronomy,
Michigan State University, East Lansing, MI 48824, USA; email: leed@frib.msu.edu}}

\begin{abstract}
Lattice effective field theory applies the principles of effective field theory in a lattice framework where space and time are discretized.  Nucleons are placed on the lattice sites, and the interactions are tuned to replicate the observed features of the nuclear force.  Monte Carlo simulations are then employed to predict the properties of nuclear few- and many-body systems.  We review the basic methods and several theoretical and algorithmic advances that have been used to further our understanding of atomic nuclei.
\end{abstract}

\maketitle

\tableofcontents

\section{Introduction}
There are many \textit{ab initio} 
approaches currently being used to study nuclear few- and many-body systems.  Some of the methods include the no-core shell model~\cite{Navratil:2003ef,Maris:2008ax,Barrett:2013nh,Wirth:2017bpw}, no-core shell model with continuum~\cite{Navratil:2016ycn,Hupin:2018biv,Hebborn:2022iiz},
symmetry-adapted no-core shell model~\cite{Dreyfuss:2020lss,Becker:2023dqe},
diffusion or Green's function Monte Carlo simulations~\cite{Carlson:2014vla,Dawkins:2019vcr,King:2022zkz},
auxiliary-field diffusion Monte Carlo simulations~\cite{Lonardoni:2018nob,Lovato:2022apd}, configuration-interaction Monte Carlo simulations~\cite{Arthuis:2022ixv}, self-consistent Green's
functions~\cite{Idini:2019hkq,Carbone:2019pkr,Barbieri:2021ezv}, many-body perturbation theory~\cite{Tichai:2020dna,Demol:2020mzd,Frosini:2021ddm}, in-medium similarity renormalization
group~\cite{Holt:2019gmc,Stroberg:2019mxo,Yao:2019rck,Hu:2021trw}, coupled cluster methods~\cite{Sun:2018fmu,Jiang:2020the,Hu:2021trw}, projected generator coordinate method with perturbation theory~\cite{Frosini:2021fjf,Frosini:2021sxj,Frosini:2021ddm}, and Monte Carlo shell model~\cite{Otsuka:2022bcf}. Lattice effective field theory is another \textit{ab initio} approach that uses the principles of effective field theory (EFT) within a lattice framework.  Protons and neutrons are placed on discrete lattice points that span space and time.  Instead of resolving the internal substructure of the nucleons as one might in lattice quantum chromodynamics, we take a lattice spacing comparable to $1$~fm or so and resolve the long-distance physics associated with low-energy nuclear systems.  Since the methodology of lattice EFT is different from that of the other approaches, it provides complementary information that may be difficult to access otherwise.  

Chiral EFT describes the low-energy physics of nucleons mediated by short-range interactions, the exchange of pions, and electromagnetic interactions.  The interactions of nucleons can be organized by counting powers of nucleon momenta along with factors of the pion mass. The terms with $n$ powers of nucleon momenta and pion mass are labeled as order $Q^n$.  At low energies, the dominant interactions are at order $Q^0$, or leading order (LO).  There are no terms at order $Q^1$, and next-to-leading order (NLO) interactions are order $Q^2$, next-to-next-to-leading order (N2LO) are $Q^3$, and 
 next-to-next-to-next-to-leading order (N3LO) are  $Q^4$. Three-nucleon forces first appear at N2LO and four-nucleon forces appear at N3LO. Some of the early work on chiral EFT can be found in Refs.~\cite{Weinberg:1990rz,Weinberg:1991um,Weinberg:1992yk,Ordonez:1992xp,Ordonez:1993tn,Friar:1994,vanKolck:1994yi,Bernard:1995dp,Ordonez:1996rz,Kaiser:1997mw,Friar:1998zt,Epelbaum:1998na,Epelbaum:1998hg,Epelbaum:1998ka}. See Refs.~\cite{Epelbaum:2008ga,Machleidt:2016rvv, Hammer:2019poc} for reviews of chiral EFT.

Ref.~\cite{Muller:1999cp} presented the first lattice calculations of nucleons with short-range interactions.  Chiral perturbation theory for pions on the lattice was studied in Refs.~\cite{Shushpanov:1998ms,Lewis:2000cc}, and chiral symmetry with static baryons were explored in Ref.~\cite{Chandrasekharan:2003wy}.  The first lattice calculations involving dynamical nucleons and chiral EFT appeared in Ref.~\cite{Lee:2004si}.  One of the challenges in this work is that the pions were also treated as dynamical fields that coupled to the nucleon fields, and this produced unwanted modifications to the properties of individual nucleons, such as the nucleon mass and the pion-nucleon coupling strength.  There were also simulations of pionless effective field theory without pions \cite{Lee:2004qd,Lee:2005is,Lee:2005it,Borasoy:2005yc}.  Pionless EFT is useful at long-distance scales, where the exchange of pions can be regarded as short-range interactions.  Several groups have applied lattice EFT to cold atoms and dilute neutron matter using pionless EFT \cite{Chen:2003vy,Bulgac:2005pj,Burovski:2006,Wingate:2006wy,Abe:2007fe,Alexandru:2019gmp,Alexandru:2020zti}.

Nuclear lattice effective field theory (NLEFT) studies using chiral EFT with both protons and neutrons began with Ref.~\cite{Borasoy:2006qn}. This work replaced the fully dynamical pion fields used in Ref.~\cite{Lee:2004si} with pion fields with only spatial correlations.  This choice produced the instantaneous nucleonic interactions that are standard in chiral EFT.  Several reviews of NLEFT and related methods can be found in Refs.~\cite{Lee:2008fa,Drut:2012md,Lee:2016fhn,Lahde:2019npb}.  One of the challenges in writing a review of NLEFT is keeping up with new developments that progress quickly and cover a wide range of topics.  In this review, we cover some of the more recent advances that have happened in the past few years and focus on new algorithms and methods that have enabled these advances.

\section{Lattice Hamiltonian}
Let us define $a_{i,j}({\bf n})$ and $a^{\dagger}_{i,j}({\bf n})$ to be the lattice
annihilation and creation operators on lattice site ${\bf n}$ with spin $i=0,1$
(up, down) and isospin $j=0,1$ (proton, neutron).   We write $a$ for the spatial lattice spacing, $a_t$ for the lattice spacing in the time direction, and $\alpha_t$ for the ratio $a_t/a$.  We set the speed of light, $c$, reduced Planck constant, $\hbar$, and Boltzmann constant, $k_B$, all equal to $1$ and work in lattice units where powers of $a$ are multiplied to make dimensionless combinations.  In most cases, the lattice is assumed to have a cubic structure with periodic boundaries.  However, other boundary conditions \cite{Korber:2015rce,Lu:2019nbg} as well as body-centered cubic lattices \cite{Song:2021yst} have also been explored. 

For the free-nucleon lattice Hamiltonian, we use a one-nucleon lattice operator that approximates the non-relativistic kinetic energy for the nucleons, $E(\vec{p}) = \vec{p}^2/2m$, where $m$ is the nucleon mass. 
 This is constructed by having the nucleons hop to various neighboring lattice sites along each lattice direction.  For any lattice direction, $l = 1,2,3$ $(\hat{\bf l} = \hat{\bf 1},\hat{\bf 2},\hat{\bf 3})$, a displacement of $k$ lattice units along that direction produces a phase factor of $e^{ikp_l}$ when applied to a nucleon with momentum $\vec{p}$.  With a linear combination of these lattice hopping operators, we can reproduce $\vec{p}^2/2m$ with an error that scales with any desired power of $p_l$ in the low-momentum limit.  For example, if we use an $O(a^4)$-improved action \cite{Lee:2008fa} with hopping terms with up to three lattice units, the error terms start at $O(p_l^8)$ and the free Hamiltonian has the form
\begin{align}
H_{\rm free}= &\frac{49}{12m}\sum_{\bf n} a^\dagger_{i,j}({\bf n}) a_{i,j}({\bf n})-\frac{3}{4m}\sum_{{\bf
n},l}
\sum_{{\bf n'}={\bf n}\pm \hat{\bf l}} a^\dagger_{i,j}({\bf n'}) a_{i,j}({\bf n}) \nonumber
\\
&+\frac{3}{40m}\sum_{{\bf
n},l}\sum_{{\bf n'}={\bf n}\pm 2\hat{\bf l}} a^\dagger_{i,j}({\bf
n'})
a_{i,j}({\bf n})-\frac{1}{180m}\sum_{{\bf
n},l}
\sum_{{\bf n'}={\bf n}\pm 3\hat{\bf l}} a^\dagger_{i,j}({\bf
n'}) a_{i,j}({\bf n}).
\end{align}
We can also use Fourier transforms into momentum space, as done in Ref.~\cite{Bulgac:2005a}, to reproduce $\vec{p}^2/2m$ exactly for all values of $p_l$ available for a given periodic lattice volume.

In a similar manner, we can use finite differences to define the operation $\nabla_{l}$ acting on a general lattice function $f(\textbf{n})$ to reproduce $ip_l$ in momentum space, with an error that scales as any desired power of $p_l$ in the low-momentum limit.  For example, if we use an $O(a^4)$-improved action, the error terms start at $O(p_l^7)$ and the form of $\nabla_l$ is 
\begin{align}
\nabla_l f({\bf n})= & \frac{3}{4}\left[f({\bf n}+{\hat{\bf l}})-f({\bf
n}-{\hat{\bf l}})\right] 
-\frac{3}{20}\left[f({\bf n}+2{\hat{\bf l}})-f({\bf
n}-2{\hat{\bf l}})\right]\nonumber \\
& +\frac{1}{60}\left[f({\bf n}+3{\hat{\bf l}})-f({\bf
n}-3{\hat{\bf l}})\right].
\end{align}
If we only need even powers of $p_l$, then we can reduce the displacements to one-half lattice units and define $\nabla_{1/2,l}$ as \cite{Li:2018ymw}
\begin{align}
\nabla_{1/2,l} f({\bf n})= & \frac{3}{2}\left[f({\bf n}+\frac{1}{2}{\hat{\bf l}})-f({\bf
n}-\frac{1}{2}{\hat{\bf l}})\right] 
-\frac{3}{10}\left[f({\bf n}+{\hat{\bf l}})-f({\bf
n}-{\hat{\bf l}})\right]\nonumber \\
& +\frac{1}{30}\left[f({\bf n}+\frac{3}{2}{\hat{\bf l}})-f({\bf
n}-\frac{3}{2}{\hat{\bf l}})\right].
\end{align}

We also define point-like density operators that depend on spin and isospin.  For spin indices $S=1,2,3,$ and isospin indices $I=1,2,3$, we let 
\begin{align}
\rho({\bf n})&= \sum_{i,j}a^\dagger_{i,j}({\bf n}) a_{i,j}({\bf n}),
\nonumber \\
\rho_{S}({\bf n})&=\sum_{i,i',j}a^\dagger_{i,j}({\bf n})[\sigma_S]_{ii'} a_{i',j}({\bf
n}),
\nonumber \\
\rho_{I}({\bf n})&=\sum_{i,j,j'}a^\dagger_{i,j}({\bf n})[\tau_I]_{jj'} a_{i,j'}({\bf
n}),
\nonumber \\
\rho_{S,I}({\bf n})&=\sum_{i,i',j,j'}a^\dagger_{i,j}({\bf n})[\sigma_S]_{ii'}
[\tau_I]_{jj'} a_{i
',j'}({\bf n}), \label{eq:densities}
\end{align}
where $\sigma_S$ are Pauli matrices in spin space  and $\tau_I$ are Pauli matrices in isospin space.  

In the following, we use the term local to describe operators where the annihilation and creation operators appear at the same location.  The term nonlocal therefore refers to operators where the annihilation and creations operators can appear at different locations. 
 It is convenient to define a local smearing parameter $s_{\rm L}$ so that the density operators in Eq.~\eqref{eq:densities} are spread over more than one lattice site,
\begin{align}
    \rho^{s_{\rm L}}({\bf n}) &= \rho({\bf n}) + s_{\rm L}\sum_{|{\bf n'}|=1}\rho({\bf
n}+{\bf n'}). \nonumber \\
    \rho^{s_{\rm L}}_S({\bf n}) &= \rho_S({\bf n}) + s_{\rm L}\sum_{|{\bf n'}|=1}\rho_S({\bf
n}+{\bf n'}). \nonumber \\
    \rho^{s_{\rm L}}_I({\bf n}) &= \rho_I({\bf n}) + s_{\rm L}\sum_{|{\bf n'}|=1}\rho_I({\bf
n}+{\bf n'}). \nonumber \\
    \rho^{s_{\rm L}}_{S,I}({\bf n}) &= \rho_{S,I}({\bf n}) + s_{\rm L}\sum_{|{\bf n'}|=1}\rho_{S,I}({\bf
n}+{\bf n'}).   
\end{align}
The local smearing can be extended beyond nearest neighbors in a straightforward manner.

With these definitions, we can develop the chiral interactions on the lattice in a straightforward manner by taking the continuum version of the chiral interactions and making the appropriate replacements of integrals with lattice sums and derivatives with lattice finite differences.  This approach was used to study dilute neutron matter at NLO \cite{Epelbaum:2009rkz}.  The first {\it ab initio} calculations of the Hoyle state of $^{12}$C were carried out using an interaction at N2LO \cite{Epelbaum:2011md,Epelbaum:2012qn}.  The spectrum and structure of the low-lying states of $^{16}$O were also calculated at N2LO \cite{Epelbaum:2013paa}.  A discussion and analysis of the interactions at N2LO are detailed in Ref.~\cite{Alarcon:2017zcv}. 

In Ref.~\cite{Li:2018ymw}, a different approach to the lattice chiral interactions was developed based on partial wave projections and nonlocal smearing functions.  The angular dependence of the relative separation between the two nucleons was prescribed by spherical harmonics, and the dependence on the nucleon spins was given by spin-orbit Clebsch-Gordan coefficients.  We define the operators $a^{s_{\rm NL}}_{i,j}({\bf n})$ and $a^{s_{\rm NL}\dagger}_{i,j}({\bf n})$ with nonlocal smearing parameter $s_{\rm NL}$,
\begin{equation}
 a^{s_{\rm NL}}_{i,j}({\bf n})=a_{i,j}({\bf n})+s_{\rm NL}\sum_{|{\bf n'}|=1}a_{i,j}({\bf
n}+{\bf n'}).
\end{equation}
\begin{equation}
 a^{s_{\rm NL}\dagger}_{i,j}({\bf n})=a^{\dagger}_{i,j}({\bf n})+s_{\rm NL}\sum_{|{\bf
n'}|=1}a^{\dagger}_{i,j}({\bf
n}+{\bf n'}).
\end{equation}
The nonlocal smearing can be extended beyond nearest neighbors in a straightforward manner.  We define the following two-by-two matrices to make a spin-$0$ combination,
\begin{equation}
M_{ii'}(0,0) = \frac{1}{\sqrt{2}}[\delta_{i,0}\delta_{i',1}-\delta_{i,1}\delta_{i',0}],
\end{equation}
and spin-$1$ combinations,
\begin{align}
& M_{ii'}(1,1) = \delta_{i,0}\delta_{i',0}, \nonumber \\
& M_{ii'}(1,0) = \frac{1}{\sqrt{2}}[\delta_{i,0}\delta_{i',1}+\delta_{i,1}\delta_{i',0}], \nonumber \\
& M_{ii'}(1,-1) = \delta_{i,1}\delta_{i',1}.
\end{align}

We can define the pair annihilation operators $[a({\bf n})a({\bf n'})]^{s_{\rm
NL}}_{S,S_z,I,I_z}$, where
\begin{equation}
 [a({\bf n})a({\bf n'})]^{s_{\rm
NL}}_{S,S_z,I,I_z}=\sum_{i,j,i',j'} a^{s_{\rm NL}}_{i,j}({\bf n})M_{ii'}(S,S_z)M_{jj'}(I,I_z)a^{s_{\rm
NL}}_{i',j'}({\bf n'}),
\label{spin-isospin}
\end{equation}
with spin quantum numbers $S,S_z$ and isospin quantum numbers $I,I_z$.
We also define the solid harmonics
\begin{equation}
R_{L,L_z}({\bf r}) = \sqrt{\frac{4\pi}{2L+1}}r^L Y_{L,L_z}(\theta,\phi),
\end{equation}
and their complex conjugates
\begin{equation}
R^*_{L,L_z}({\bf r}) = \sqrt{\frac{4\pi}{2L+1}}r^L Y^*_{L,L_z}(\theta,\phi).
\end{equation}
We note that $R_{L,L_z}$ and $R^*_{L,L_z}$ are homogeneous polynomials with degree $L$. 

Using the pair annihilation operators, lattice finite differences,
and solid harmonics, we form the operator combinations
\begin{equation}
P^{2M,s_{\rm NL}}_{S,S_z,L,L_z,I,I_z}({\bf n}) =[a({\bf n}){\bf \nabla}^{2M}_{1/2}R^*_{
L,L_{z}}({\bf \nabla})a({\bf n})]^{s_{\rm
NL}}_{S,S_z,I,I_z}, 
\end{equation}
where ${\bf \nabla}_{1/2}^{2M}$ and ${\bf \nabla}$ act on the second
annihilation operator.  This means we act on ${\bf n'}$ in Eq.~(\ref{spin-isospin}) and then set ${\bf n'}$ to equal
${\bf n}$.  The even integer $2M$ introduces extra derivatives. Writing the Clebsch-Gordan coefficients as $\langle S S_z, L L_z \vert J J_z\rangle$,
we define 
\begin{equation}
O^{2M,s_{NL}}_{S,L,J,J_z,I,I_z}({\bf n})= 
\sum_{S_z,L_z}\langle S S_z, L L_z \vert J J_z\rangle P^{2M,s_{NL}}_{S,S_z,L,L_z,I,I_z}({\bf
n}).
\end{equation}
Using $O^{2M,s_{NL}}_{S,L,J,J_z,I,I_z}({\bf n})$ and its Hermitian conjugate, $[O^{2M,s_{NL}}_{S,L,J,J_z,I,I_z}({\bf n})]^\dagger$, we can construct short-range operators two-nucleon operators up to any order.  This was done for operators up to N3LO in \cite{Li:2018ymw}.  If we split apart $O^{2M,s_{NL}}_{S,L,J,J_z,I,I_z}({\bf n})$ and $[O^{2M,s_{NL}}_{S,L,J,J_z,I,I_z}({\bf n}')]^\dagger$ so that they are located on neighboring lattice sites, then we can introduce a dependence on the total momentum $\textbf{P}$ of the two nucleons of the form $e^{i\textbf{P}\cdot(\textbf{n}-\textbf{n}')}$.  This was used in Ref.~\cite{Li:2019ldq} to restore Galilean invariance on the lattice, order by order in powers of momentum.

\section{Lattice scattering with spherical wall boundaries}
Lüscher's finite volume method \cite{Luscher:1986pf,Luscher:1990ck,Luscher:1990ux} is commonly used to determine phase shifts on the lattice from finite-volume energies.  However, one must contend with the cubic periodic boundaries introducing rotational symmetry breaking effects that go beyond the rotational symmetry breaking due to the lattice spacing.  This is something one likely cannot avoid if the scattering bodies are composite objects computed from many-body simulations.  For the simpler case of two-nucleon scattering, however, one can obtain much better performance by using spherical wall boundaries to remove the effect of the periodic box.  

The basic idea is to impose a spherical wall boundary on the relative separation between the two nucleons and analyze the spectrum and wavefunctions of the standing waves that result.  This technique was used in continuous space calculations in Ref.~\cite{Carlson:1984zz} and implemented on the lattice in Ref.~\cite{Borasoy:2007vy}. 
 In Ref.~\cite{Lu:2015riz}, the technique was very significantly improved by using spherical harmonics to project the three-dimensional lattice into radial bins with specified orbital angular momentum quantum numbers.  The calculation of the mixing angles for the coupled partial waves was also greatly improved by introducing a complex-valued auxiliary potential at large radial distances just inside the spherical wall boundary.  The purpose of this auxiliary potential was to break time reversal symmetry so that the real and imaginary parts of the standing wave solutions were linearly independent, thereby providing the necessary information to extract the mixing angle at the energy of the standing wave.  This technique was generalized to work for an arbitrary number of coupled channels in Ref.~\cite{Bovermann:2019jbt}.

\begin{figure}
\centering
\includegraphics[width=\textwidth]{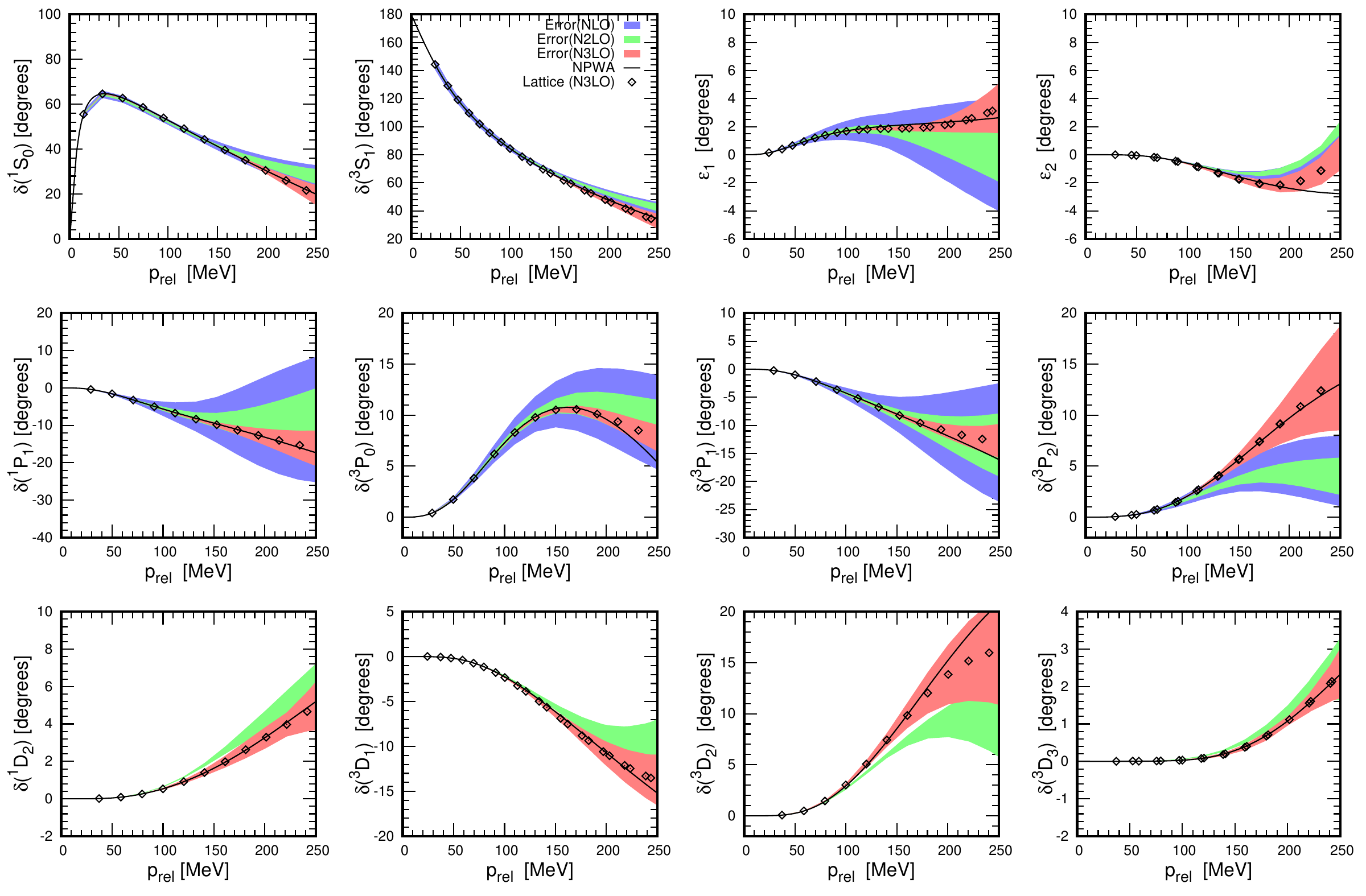}
\caption{Neutron-proton scattering phase shifts and mixing angles with theoretical uncertainties versus the relative momenta 
for $a = 1.32~{\rm fm}$.  Blue, green, and red bands signify the estimated uncertainties at NLO, N2LO, and N3LO, respectively. The black solid line and diamonds 
denote the phase shift or mixing angle from the Nijmegen partial-wave analysis \cite{Stoks:1993tb} and lattice calculation at N3LO, respectively. Figure taken with permission from Ref.~\cite{Li:2018ymw}.}
\label{Errora150}
\end{figure}

In Fig.~\ref{Errora150}, we show phase shifts and mixing angles at $O(Q^4)$ or N3LO for lattice spacing $a=1.32$~fm \cite{Li:2018ymw}.  We plot results 
for relative momenta up to $p_{\rm rel} = 250~{\rm MeV}$.  The theoretical uncertainties were calculated using the method described in Refs.~\cite{Epelbaum:2014efa,Epelbaum:2014sza} that estimates the convergence of the chiral expansion and predicts the remaining error from higher-order terms.  Other approaches for estimating theoretical uncertainties have also been discussed in Refs.~\cite{Furnstahl:2014xsa,Furnstahl:2015rha}.

\section{Euclidean-time projection and perturbation theory}
The Euclidean-time transfer matrix
$M$ is defined as the normal-ordered exponential of the lattice Hamiltonian
$H$ over one time lattice step,
\begin{equation}
M = :\exp(-H\alpha_t):.
\label{eq:transfer_matrix}
\end{equation}
Normal ordering, as indicated in Eq.~\eqref{eq:transfer_matrix} by the $::$ symbols, corresponds to placing the annihilation operators to the right and the creation operators to the left and multiplying by the necessary minus signs when reordering operators.

We consider Euclidean-time evolution starting from an initial state corresponding to a Slater determinant of single-nucleon wavefunctions, $\ket{\Psi_{\rm init}}$.  By multiplying many powers of $M$ upon $\ket{\Psi_{\rm init}}$, we project out the lowest-energy state that has nonzero overlap with $\ket{\Psi_{\rm init}}$.  We compute projection amplitudes of the form
\begin{equation}
Z(n_t) = \braket{\Psi_{\rm init}| M^{n_t}  |\Psi_{\rm init}}, \label{eq:amp}
\end{equation}
and also compute matrix elements of observables $O$, 
\begin{equation}
Z_O(n_t) = \braket{\Psi_{\rm init}| M^{n_t/2} O M^{n_t/2}  |\Psi_{\rm init}}. \label{eq:expect}
\end{equation}
We have assumed that $n_t$ is even.

There are infinitely many interactions in chiral EFT, and they are organized according to their expected importance at low energies.  It is therefore very helpful to use perturbation theory when performing calculations.  Some set of interactions that are dominant at low energies are treated non-perturbatively, e.g., the LO interactions, and the rest are treated using perturbation theory.  We illustrate with a simple example.  Suppose that the Hamiltonian has the form
\begin{equation}
    H(c) = H_0 + c H_1.
\end{equation}
We can expand the normal-ordered transfer matrix to first order in $c$,
\begin{equation}
M(c) = M + c M' + O(c^2).
\end{equation}
When considered as a function $c$, we let $Z'(n_t)$ be the first-order correction to $Z(n_t)$ and let $Z'_O(n_t)$ be the first-order correction to $Z_O(n_t)$.  $Z'(n_t)$ corresponds to replacing one of the $n_t$ insertions of $M$ by $M'$ in Eq.~\ref{eq:amp} and  $Z'_O(n_t)$ corresponds to replacing one of the $n_t$ insertions of $M$ by $M'$ in Eq.~\ref{eq:expect}. 

We can go beyond first-order perturbation theory by allowing for more than one insertion of $M'$.  In order to perform higher-order perturbation theory calculations, one must take care that stochastic noise fluctuations do not become too large.  The first NLEFT calculations using second-order perturbation theory for computing energies were performed in Ref.~\cite{Lu:2021tab}.  One of the techniques used in Ref.~\cite{Lu:2021tab} to reduce statistical noise was to shift the integration contour in the complex plane for the auxiliary field.  Second-order perturbation theory for computing energies in continuous-space quantum Monte Carlo simulations is also discussed in Ref.~\cite{Curry:2024gcz}.  For multichannel calculations, we take several different Slater-determinant initial states $\ket{\Psi_{{\rm init},1}}, \ket{\Psi_{{\rm init},2}}, \cdots$
and compute a matrix of amplitudes and operator matrix elements using these initial states. 

\section{Auxiliary-field Monte Carlo simulations}
It is convenient to use auxiliary fields to generate the short-range nuclear interactions.  The basic idea is to rewrite the nucleonic interactions as couplings between nucleons and fluctuating auxiliary fields.  We illustrate with a simple example. Let $\rho({\bf n})$ be the total nucleon density at some particular lattice site ${\bf n}$.  The exact Gaussian integral identity,
\begin{equation}
{:\exp\left(-\frac{c\alpha_t}{2}\rho^2({\bf n})\right):}=\sqrt{\frac{1}{2\pi}}\int^{\infty}_{-\infty}ds({\bf n})
\, {:\exp \left(-\frac{1}{2}s^2({\bf n}) + \sqrt{-c\alpha_t}s({\bf n})\rho({\bf n}) \right):}\;, \label{HS}
\end{equation}
allows us to replace the two-nucleon interaction by a one-nucleon interaction with the auxiliary field $s({\bf n})$.  This transformation can be performed at every lattice site.  

\begin{figure}[h]
\centering
\includegraphics[width=0.95\columnwidth,trim={0.5cm 0.5cm 0.5cm 0.5cm},clip]{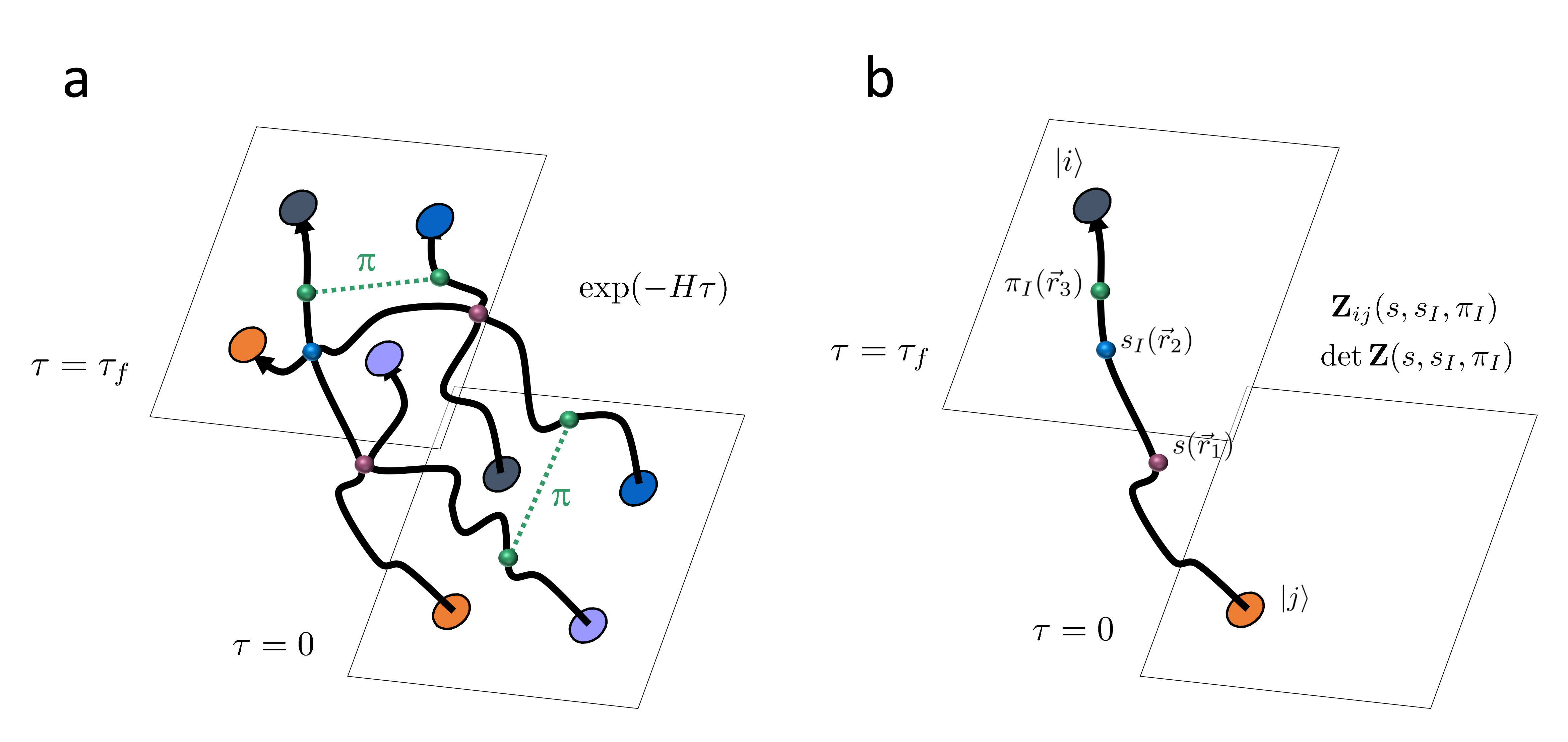}
\caption{Panel {\bf a} shows the Euclidean-time evolution using projection Monte Carlo.  Panel {\bf b} shows the interactions of individual nucleons with auxiliary fields and pion fields.}
\label{fig:Euclidean_time}
\end{figure}

In this manner, we introduce auxiliary fields to produce the required two-particle interactions \cite{Hubbard:1959ub,Stratonovich:1958,Koonin:1986}.  Higher-body interactions can also be produced using auxiliary fields \cite{Chen:2004rq,Korber:2017emn}. 
The pion fields that produce instantaneous pion-exchange potentials can also be viewed as being analogous to auxiliary fields.  However, the pion fields also have nontrivial correlations across spatial lattice sites. In the auxiliary-field formalism, the many-body amplitude for an $A$-body system corresponds to the determinant of an $A \times A$ matrix of single-nucleon amplitudes.  This is illustrated in Fig.~\ref{fig:Euclidean_time}. Panel {\bf a} shows the Euclidean-time evolution using projection Monte Carlo.  Panel {\bf b} shows the interactions of individual nucleons with auxiliary fields and pion fields.  One needs to sample over all configurations for the auxiliary and pion fields.  This is performed using Markov chain Monte Carlo sampling.  In Ref.~\cite{Lu:2018bat}, a technique called the shuttle algorithm was introduced for efficient updates of the auxiliary fields and pion fields while systematically traversing the lattice forward and backward in Euclidean time.

It is important that the matrix determinant does not fluctuate strongly in sign or complex phase.  Otherwise, the calculation will suffer from the well-known Monte Carlo sign problem.  Fortunately, the dominant nucleon-nucleon interactions at low energies are two S-wave interactions in the $^1{\rm S}_0$ and $^3{\rm S}_1$ channels that are both attractive and approximately equal in strength.  This is often described as Wigner's approximate SU(4) symmetry \cite{Wigner:1936dx} where the four nucleonic degrees of freedom (proton spin-up, proton spin-down, neutron spin-up, neutron spin-down) can be viewed as four components of an SU(4) multiplet. There are some interesting connections between Wigner's SU(4) symmetry and quantum chromodynamics with many colors \cite{Kaplan:1995yg,Kaplan:1996rk,Mehen:1999qs,Banerjee:2001js,CalleCordon:2008cz,Lee:2020esp}. In the limit of exact SU(4) symmetry with attractive interactions, the $A \times A$ matrix of single-nucleon amplitudes has a block-diagonal structure with one block each for proton spin-up, proton spin-down, neutron spin-up, and neutron spin-down.  The determinant of such a matrix is positive semidefinite.  The protection against the sign problem provided by SU(4) symmetry is discussed in Refs.~\cite{Chen:2004rq,Lee:2007eu}.  In the absence of sign oscillations, one can also prove rigorous inequalities, as demonstrated in Refs.~\cite{Lee:2004ze,Lee:2004hc,Chen:2004rq,Lee:2007eu,Hoffman:2016jqv}.

Monte Carlo simulations of higher-order terms in chiral EFT typically result in severe sign problems.  This is due to strong repulsive interactions at short distances.  Some efforts were made to mitigate this problem using extrapolation methods \cite{Lahde:2015ona}, and variational approaches have also been used \cite{Wlazlowski:2014jna}.  An approach called eigenvector continuation was also considered \cite{Frame:2017fah}, and we will return to the discussion of eigenvector continuation when discussing the floating block method.  Most efforts to avoid sign problems in NLEFT simulations have focused on the use of perturbation theory to implement higher-order terms in chiral EFT \cite{Borasoy:2007vk,Epelbaum:2009rkz,Epelbaum:2009zsa,Epelbaum:2009pd,Epelbaum:2010xt,Lahde:2013uqa,Li:2018ymw}.  We will revisit this point when discussing the acceleration of perturbation using wavefunction matching.

\section{Wavefunction matching for accelerated perturbation theory convergence}

When there is no sign problem, quantum Monte Carlo simulations are a very powerful {\it ab initio} method for computing quantum many-body systems.  The computational effort scales as a low power of the number of nucleons.  For NLEFT quantum Monte Carlo simulations using auxiliary fields, the scaling is approximately quadratic in the number of nucleons for medium-mass nuclei, and this slowly crosses over to cubic scaling for heavy nuclei.  In many cases, a simple Hamiltonian, $H^S$, can be found that is easily computable using quantum Monte Carlo methods and describes the energies and other properties of the many-body system in decent agreement with empirical data \cite{Lu:2018bat,Lu:2019nbg,Shen:2022bak,Gnech:2023prs}. 

An example of such a simple Hamiltonian, $H^S$, was presented in Ref.~\cite{Lu:2018bat}.  A minimal nuclear lattice Hamiltonian was introduced that had no Monte Carlo sign problem due to Wigner's SU(4) symmetry and contained only four adjustable parameters.  The Coulomb interaction was taken into account using perturbation theory.  In Ref.~\cite{Lu:2018bat}, it was shown that the minimal lattice Hamiltonian reproduces the energies and charge radii of most light and medium-mass nuclei quite well. Despite the simple form, one can also obtain a fairly good description of the carbon isotopes \cite{Shen:2022arg,Shen:2022bak}, beryllium isotopes \cite{Shen:2024qzi}, and monopole transition of the alpha particle \cite{Meissner:2023cvo}.

While simplified interactions are useful, a primary goal of {\it ab initio} nuclear theory is to describe nuclear physics with a systematic program for reducing and eliminating all sources of error.  Unfortunately, realistic high-fidelity Hamiltonians typically produce severe sign problems that make quantum Monte Carlo calculations difficult.  In Ref.~\cite{Elhatisari:2022zrb}, a new approach called wavefunction matching addresses this problem.  Wavefunction matching transforms the interaction between particles so that the low-energy wavefunctions up to some distance $R$ match that of an easily computable simple Hamiltonian, $H^S$.  This produces a rapidly converging expansion in powers of the difference $H'-H^S$ and allows for calculations of systems that would otherwise be impossible.

\begin{figure}[h]
\centering
\includegraphics[width=0.7
\columnwidth,trim={0.2cm 0.2cm 0.2cm 0.2cm},clip]{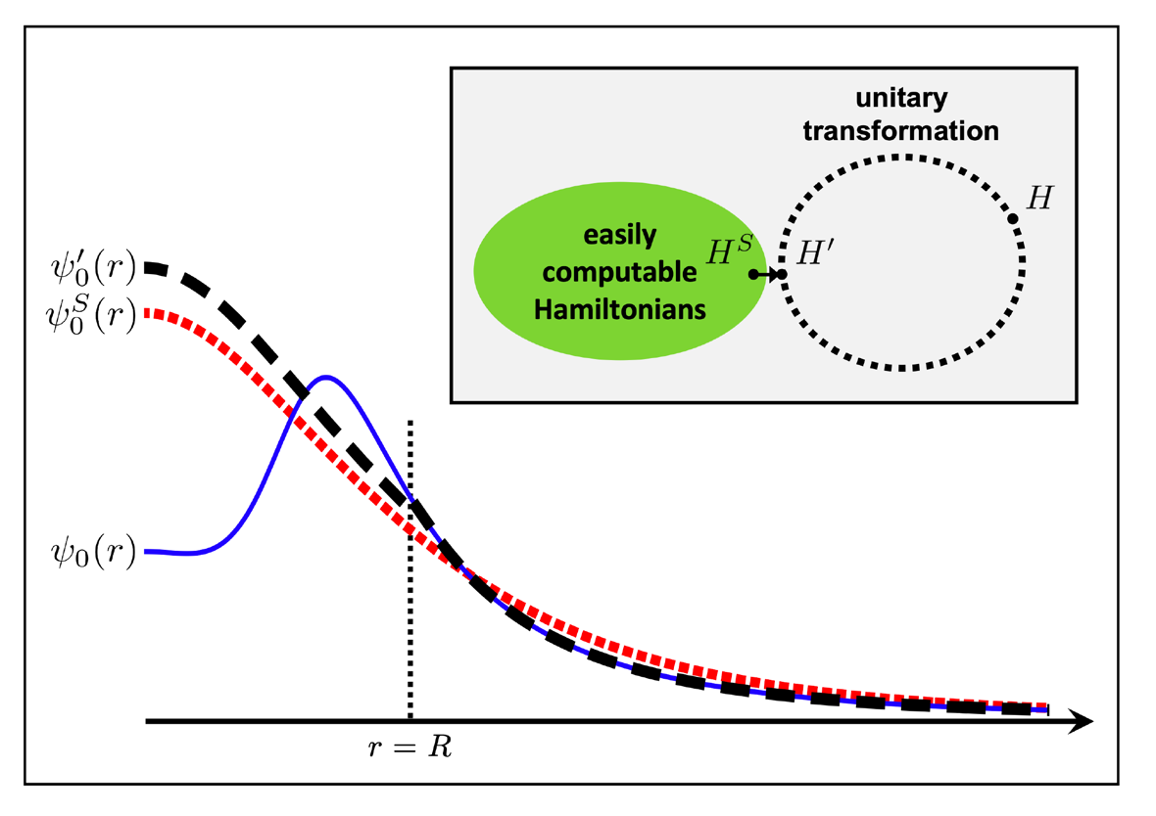}
\caption{A unitary transformation is used produce a new Hamiltonian $H'$ that is close to $H^S$. In each two-body channel, the ground state wavefunction of $H'$ matches the ground state
wavefunction of $H$ for $r > R$ and is proportional to the ground state wavefunction of $H^S$ for $r < R$, with constant of proportionality close to $1$. Figure taken with permission from Ref.~\cite{Elhatisari:2022zrb}.}
\label{fig:wfm}
\end{figure}

In Fig.~\ref{fig:wfm}, we show the scattering wavefunctions $\psi_0$ for the original high-fidelity Hamiltonian, $H$, and $\psi^S_0$ for the simple Hamiltonian, $H^S$. A unitary transformation with finite range $R$ is used to replace the short distance part of $\psi_0$ with the short distance part of $\psi^S_0$, up to a normalization factor close to $1$.  The result defines the new wavefunction $\psi'_0$, and the same unitary transformation also defines the transformed high-fidelity Hamiltonian $H'$.  The similarity between the wavefunctions $\psi'_0$ and $\psi^S_0$ is responsible for the accelerated convergence of perturbation theory.  In Ref.~\cite{Elhatisari:2022zrb}, wavefunction matching was used to accurately determine the binding energies of the light and medium-mass nuclei, charge radii of light and medium-mass nuclei, and ground state energies of symmetric nuclear matter and neutron matter.  The Supplementary Information given in Ref.~\cite{Elhatisari:2022zrb} shows how wavefunction matching can also be used for continuous space calculations.

N3LO chiral EFT interactions with wavefunction matching have been used in NLEFT calculations of the spin and density correlations in hot neutron matter \cite{Ma:2023ahg}, charge radii of silicon isotopes \cite{Konig:2023rwe}, spectrum and structure of beryllium isotopes \cite{Shen:2024qzi}, structure of ${}^{22}$Si \cite{Zhang:2024wfd}, and triton lifetime \cite{Elhatisari:2024otn}.

\section{Pinhole algorithm for $A$-body densities}

When Euclidean-time projection is applied to the initial state, $\ket{\Psi_{\rm init}}$, the result is a quantum state that is a distribution of center-of-mass positions that depend on the auxiliary and pion field configurations.  This makes it difficult to describe the structure of the nucleus relative to its center of mass.  To locate the center of mass of the nucleus, one needs to measure the locations of all $A$ nucleons, and the measurement of an $A$-body operator will typically generate large statistical noise.  In Ref.~\cite{Elhatisari:2017eno}, this problem was addressed using a method called the pinhole algorithm.  The pinhole algorithm computes an $A$-body density distribution and, most importantly, performs Monte Carlo importance sampling according to the $A$-body density.  The importance sampling removes the problems of large statistical noise.

\begin{figure}[h]
\centering
\includegraphics[width=0.95
\columnwidth,trim={0.5cm 0.5cm 0.5cm 0.5cm},clip]{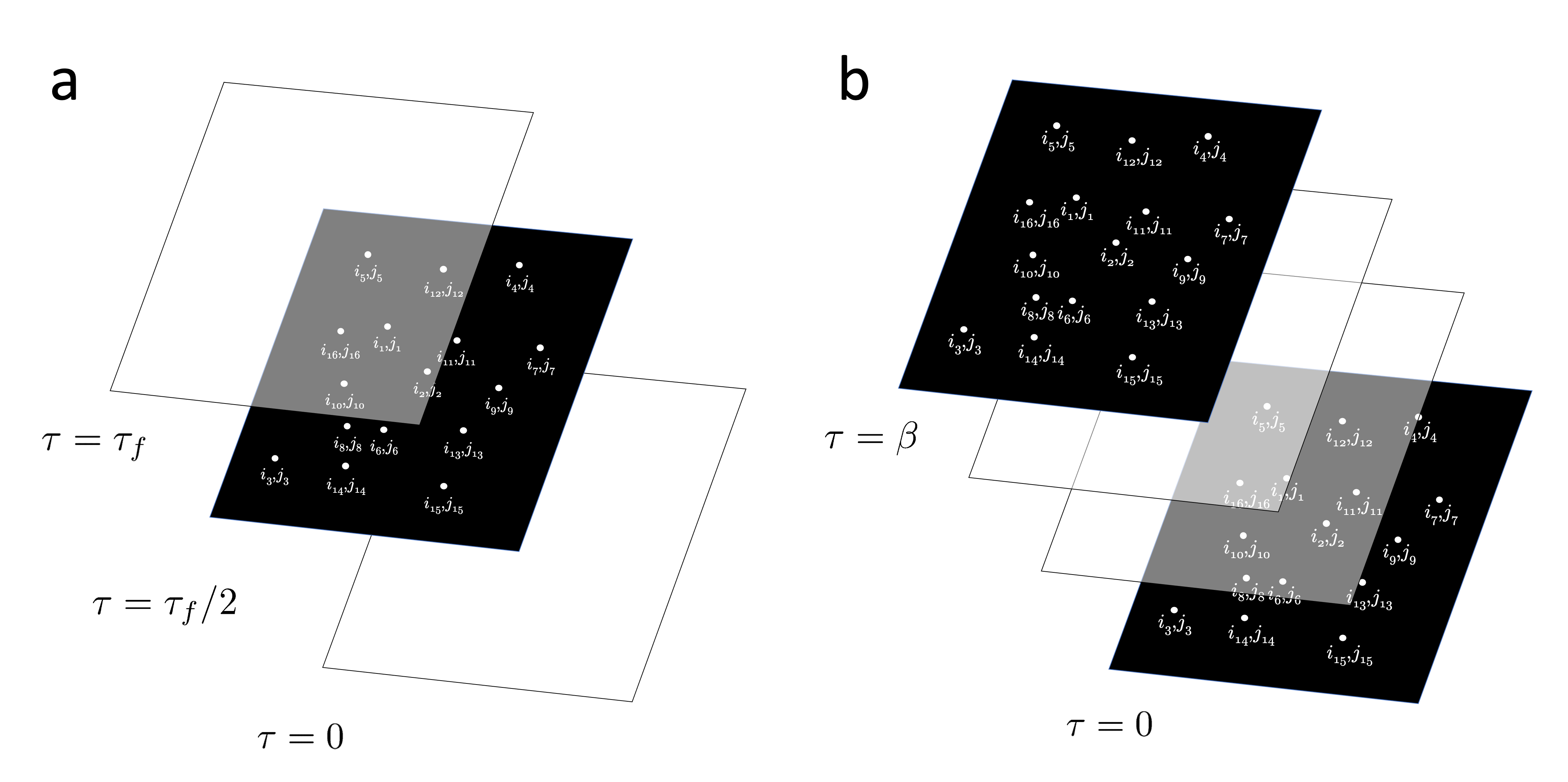}
\caption{Panel {\bf a} shows the pinhole algorithm with pinholes inserted in the midpoint of the Euclidean time evolution.  Panel {\bf b} shows the pinhole trace algorithm with the same pinholes inserted at the beginning and end of the Euclidean time evolution for the partition function, ${\rm Tr} \exp(-\beta H)$.}
\label{fig:pinholes}
\end{figure}

Let us write $\rho_{i,j}({\bf n})$ for the nucleon density operator with spin
$i$ and isospin $j$ at lattice site {\bf n},
\begin{equation}
\rho_{i,j}({\bf n}) = a^\dagger_{i,j}({\bf n})a_{i,j}({\bf n}).
\end{equation}
The normal-ordered $A$-body density operator is defined as
\begin{equation}
\rho_{i_1,j_1,\cdots i_A,j_A}({\bf n}_1,\cdots {\bf n}_A) = \; :\rho_{i_1,j_1}({\bf
n}_1)\cdots\rho_{i_A,j_A}({\bf
n}_A):.
\end{equation}
As shown in panel {\bf a} of Fig.~\ref{fig:pinholes}, we insert this $A$-body density operator at the midpoint of the Euclidean-time evolution.  The many-body amplitude vanishes unless there are $A$ nucleons that exactly match the spatial positions ${\bf n}_1,\cdots {\bf n}_A$ with spin and isospin indices $i_1,j_1,\cdots i_A,j_A$.   The spatial positions and spin-isospin indices can be viewed as ``pinholes'' in a wall that is otherwise impenetrable.  The pinhole positions and indices are updated using Monte Carlo sampling. 
When we sum over all possible spatial positions and spin-isospin indices, we get the completeness relation
\begin{equation}
\sum_{i_1,j_1,\cdots i_A,j_A}\sum_{{\bf n}_1,\cdots {\bf n}_A} \rho_{i_1,j_1,\cdots i_A,j_A}({\bf n}_1,\cdots {\bf n}_A) = A!,
\end{equation}
when acting on any $A$-body quantum state.

\begin{figure}[h]
\centering
\includegraphics[width=0.95
\columnwidth]{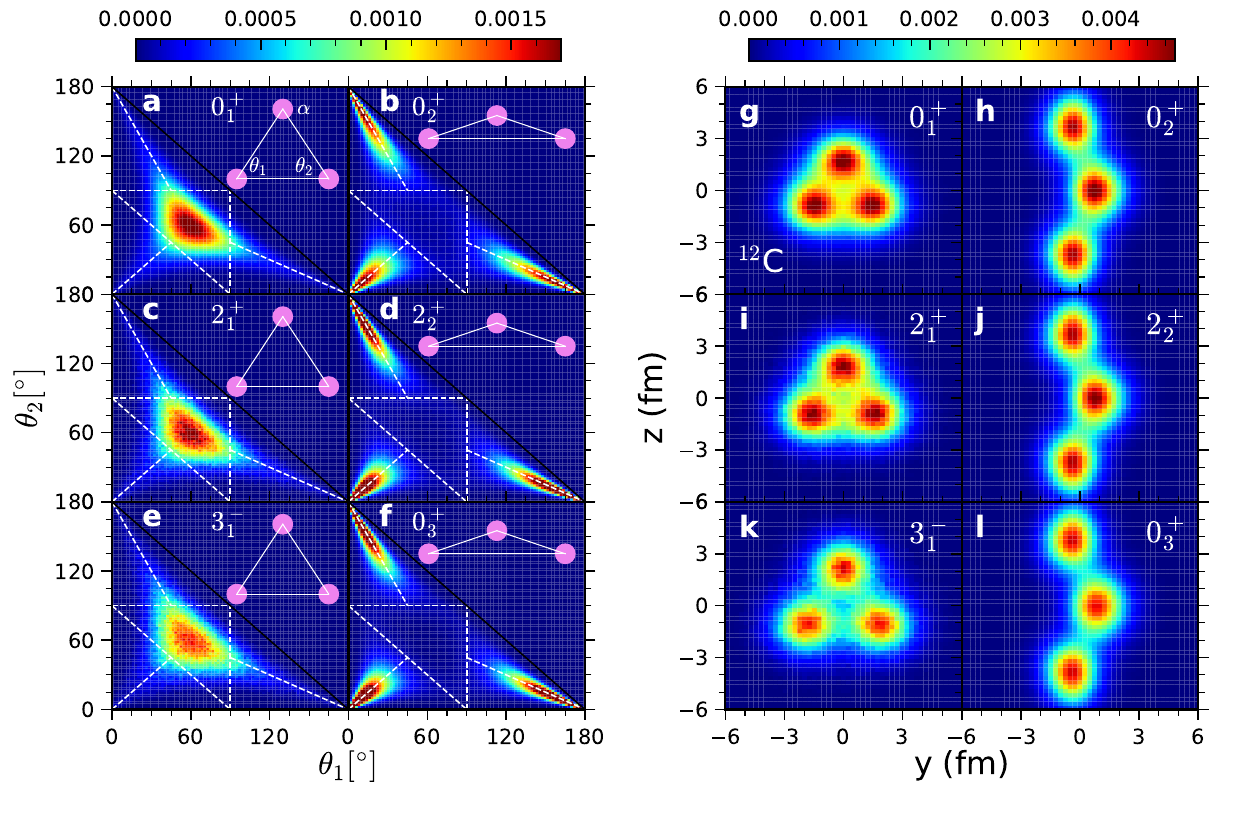}
\caption{Intrinsic nucleon densities for several low-lying states of $^{12}$C.  Panels {\bf a} through {\bf f} show the angles of the triangle formed by the three alpha clusters.  Panels {\bf g} through {\bf l} show the intrinsic densities.  Figure taken with permission from Ref.~\cite{Shen:2022bak}.}
\label{fig:densities_12C}
\end{figure}

The pinhole algorithm has been used to compute the density distributions of protons and neutrons in various nuclei \cite{Elhatisari:2017eno,Lu:2018bat}.
In Ref.~\cite{Shen:2022bak}, the pinhole algorithm was used to study the intrinsic nucleon densities for several low-lying states of $^{12}$C, improving upon a previous study \cite{Shen:2021kqr}.  Panels {\bf a} through {\bf f} of Fig.~\ref{fig:densities_12C} show the angles of the triangle formed by the three alpha clusters comprising several low-energy states in $^{12}$C.  Panels {\bf g} through {\bf l} show the corresponding intrinsic nucleon densities.  The $0^+_1$ state is the ground state of $^{12}$C, and the $0^+_2$ state is the Hoyle state. 

The pinhole configurations obtained from the NLEFT simulations are useful as initial states for relativistic ion collisions.  In ultra-relativistic collisions, the two colliding nuclei are Lorentz contracted along the beam direction, and each collision event can be viewed as a nearly simultaneous quantum measurement of the positions of the nucleons struck in the collision, before they are converted into high-energy partons.  The correlations due to deformed intrinsic shapes and clustering in the initial nuclei are captured by the pinhole configurations obtained from NLEFT simulations.  Pinhole configurations for $^{16}$O and $^{20}$Ne have been used for studies of relativistic ion collisions in Refs.~\cite{Summerfield:2021oex,Ding:2023ibq,Zhao:2024feh,Zhang:2024vkh,Giacalone:2024luz,Giacalone:2024ixe}.

\section{Pinhole trace algorithm for thermodynamics}
We can perform nuclear thermodynamics using NLEFT simulations if we take the duration of Euclidean-time evolution to be the inverse temperature, $\beta$, and impose antiperiodic boundary conditions in the time direction for the nucleons.  The antiperiodic boundary conditions for the nucleons are needed for the trace over quantum states when computing the partition function, ${\rm Tr} \exp(-\beta H)$.

A modified version of the pinhole algorithm, called the pinhole trace algorithm, was developed in Ref.~\cite{Lu:2019nbg} for first-principles calculations of nuclear thermodynamics.  In the pinhole trace algorithm, the trace over $A$-nucleon states is computed in position space,    
\begin{equation}
{\rm Tr}\; O = 
\frac{1}{A!} \sum_{i_1,j_1,{\bf n}_1 \cdots i_A,j_A,{\bf n}_A}
\langle 0 |
a_{i_A,j_A}({\bf n}_A) \cdots
a_{i_1,j_1}({\bf n}_1) \, 
O
\,
 a^{\dagger}_{i_1,j_1}({\bf n}_1) \cdots
a^{\dagger}_{i_A,j_A}({\bf n}_A)
| 0 \rangle.
\end{equation}
This is illustrated in panel {\bf b} of Fig.~\ref{fig:pinholes} for the partition function, ${\rm Tr} \exp(-\beta H)$.  

\begin{figure}[h]
\centering
\includegraphics[width=0.95
\columnwidth,trim={0.5cm 0.5cm 0.5cm 0.5cm},clip]{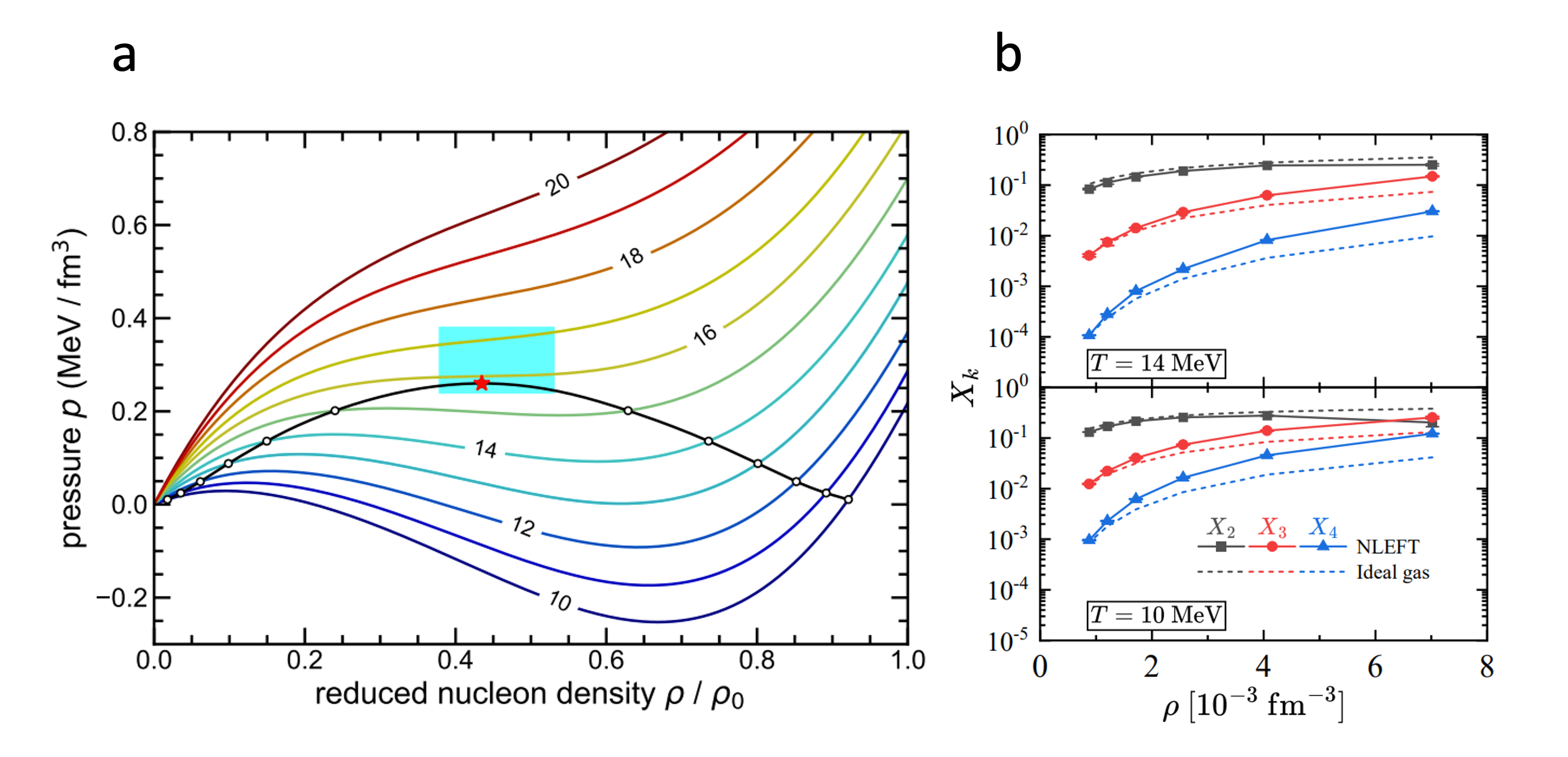}
\caption{Panel {\bf a} shows the isothermal lines of symmetric nuclear matter at the temperatures indicated in units of MeV. The black line denotes the liquid-vapor coexistence line, and the red star marks
the calculated critical point. The cyan rectangle marks the empirical critical point extracted from heavy-ion collisions \cite{Elliott:2013pna}.  Figure taken with permission from Ref.~\cite{Lu:2018bat}. Panel {\bf b} shows the abundances versus density for two-, three-, and four-nucleon clusters from lattice simulations in comparison with ideal gas results.  Figure taken with permission from Ref.~\cite{Ren:2023ued}.}
\label{fig:thermo}
\end{figure}

In Ref.~\cite{Lu:2019nbg}, the pinhole trace algorithm was introduced to perform {\it ab initio} calculations of the thermodynamics of symmetric nuclear matter.  As shown in panel {\bf a} of Fig.~\ref{fig:thermo}, the equation of state and the location of the liquid-vapor critical point were determined for the simple Hamiltonian described in Ref.~\cite{Lu:2018bat}.  Isothermal lines are shown as well as the liquid-vapor coexistence line for a plot of pressure versus reduced density $\rho/\rho_0$.  Here, $\rho_0$ is the saturation density of symmetric nuclear matter, which is about $0.16~{\rm fm}^{-3}$. We note the good agreement with the empirical critical point extracted from heavy-ion collisions \cite{Elliott:2013pna}.  

In Ref.~\cite{Ren:2023ued}, the pinhole trace algorithm and a method called light-cluster distillation was used to determine the abundance of light nuclear clusters with $A=2,3,4$ nucleons in hot and dilute symmetric nuclear matter.  Light-cluster distillation consists of measuring a set of nucleonic density correlation functions for bound light clusters with $A=2,3,4$.  The same correlation functions are then evaluated for the many-body system of interest.  Since the spatial correlations for two different clusters in the hot and dilute gaseous environment are numerically small, we can see signatures of the individual clusters in the spatial correlations and determine the abundances of the light clusters.  Panel {\bf b} of Fig.~\ref{fig:thermo} shows the abundances of two-, three-, and four-nucleon clusters versus density in comparison with ideal gas results.

\section{Adiabatic projection method for nucleus-nucleus scattering}

The adiabatic projection method builds a low-energy effective theory for clusters of nucleons \cite{Pine:2013zja,Rokash:2015hra,Elhatisari:2015iga,Elhatisari:2016hby,Elhatisari:2019fvk,Elhatisari:2021eyg}.  The method uses Euclidean-time propagation, and the low-energy effective theory becomes exact in the limit of large Euclidean time.  We start with two clusters of localized particles centered at the locations ${\bf r}$ and ${\bf r}+{\bf R}$.  We write the product of wave packets as $\ket{{\bf r}+{\bf R}}_1\otimes \ket{{\bf r}}_2$.  We sum over all values for ${\bf r}$ in our periodic volume, thereby setting the total momentum to zero.  The result is a set of two-cluster states indexed by their spatial separation vector, ${\bf R}$, 
\begin{equation}
|{\bf R}\rangle=\sum_{{\bf r}} \ket{{\bf r}+{\bf R}}_1\otimes \ket{{\bf r}}_2.
\label{eqn:single_clusters}
\end{equation}

The two-cluster states $\ket{\bf R}$ are now binned together
according to radial distance and angular momentum. This produces radial position states with quantum numbers of angular momentum, which we label as $|R\rangle^{J,J_z}$.  We multiply by powers of the transfer matrix in order to form ``dressed'' cluster
states that approximately span the set of low-energy cluster-cluster scattering
states in the periodic box. After $n_t$ time steps, the dressed cluster states are given by
\begin{equation}
\vert R\rangle^{J,J_z}_{n_t} = M^{n_t}|R\rangle^{J,J_z}.
\end{equation}
We use the dressed cluster states to compute matrix
elements of the transfer matrix $M$,
\begin{equation}
\left[M_{n_t}\right]^{J,J_z}_{R',R} =\ ^{J,J_z}_{\!\!\quad{n_t}}\langle
R'\vert M \vert R\rangle^{J,J_z}_{n_t}.
\label{Hmatrix}
\end{equation}
Since such states are not orthogonal, it is also necessary to compute the Gram or norm matrix,
\begin{equation}
\left[N_{n_t}\right]^{J,J_z}_{R',R} =\ ^{J,J_z}_{\!\!\quad{n_t}}\langle
R'\vert R\rangle^{J,J_z}_{n_t}. 
\label{eqn:norm}
\end{equation}
The ``radial adiabatic transfer matrix'' is defined as the matrix product
\begin{equation}
\left[ {M^a_{n_t}} \right]^{J,J_z}_{R',R} = 
\left[N_{n_t}^{-\frac{1}{2}}M_{n_t}
N_{n_t}^{-\frac{1}{2}} \right]^{J,J_z}_{R',R}.
\label{eqn:Adiabatic-Hamiltonian}
\end{equation}
As we did for the two-nucleon system, we use spherical wall boundaries to determine the scattering properties of the two-cluster system. The standing waves of the radial adiabatic transfer matrix with spherical wall boundaries are used to determine the elastic phase shifts.  This approach was used in Ref.~\cite{Elhatisari:2015iga} to compute the elastic scattering between two alpha particles.  By including additional channels, the radial adiabatic transfer matrix can be used to calculate inelastic reactions as well as capture reactions \cite{Rupak:2013aue}.

\section{Floating block method for eigenvector continuation}

Eigenvector continuation \cite{Frame:2017fah,Frame:2019jsw,Demol:2019yjt,Konig:2019adq,Ekstrom:2019lss,Furnstahl:2020abp} is a subspace projection method for solving parameterized eigenvalue problems.  As an example, let us consider a family of Hamiltonians, $H(c) = H_0 + c H_1,$ and some particular eigenvector, $H(c)\ket{\psi(c)} = E(c)\ket{\psi(c)}$.  The approach of eigenvector continuation is to take several snapshots of the eigenvector at points $\left\{c_m\right\}_{m=1}^N$, project the Hamiltonian onto the subspace spanned by the eigenvector snapshots, $\left\{\ket{\psi(c_m)}\right\}_{m=1}^N$, and solve the resulting generalized eigenvalue problem.  Eigenvector continuation belongs to class techniques called reduced basis methods \cite{Bonilla:2022rph,Melendez:2022kid,hesthaven2015certified,Quarteroni:218966}. A review of eigenvector continuation can be found in Ref.~\cite{Duguet:2023wuh}.
 
 In order to solve the generalized eigenvalue problem, we need to compute inner products between energy eigenstates associated with different Hamiltonians.  However, calculating the inner products of eigenstates of different Hamiltonians is not straightforward using Monte Carlo simulations and Euclidean-time projection.  We illustrate with two Hamiltonians $H_A$ and $H_B$ with ground state wavefunctions $\ket{v^0_A}$ and $\ket{v^0_B}$, respectively, and ground state energies $E^0_A$ and $E^0_B$, respectively.  Let $\ket{\phi}$ be any state with nonzero overlap with both $\ket{v^0_A}$ and $\ket{v^0_B}$.  Starting from the state $\ket{\phi}$, we can obtain $\ket{v^0_A}$ by applying $e^{-H_A t}$ for large and positive $t$.  Similarly, we can obtain $\ket{v^0_B}$ by applying $e^{-H_B t}$ for large and positive $t$.  In the limit of large $t$, we get
\begin{gather}
    e^{-H_A t} \ket{\phi} \approx e^{-E^0_A t} \braket{v^0_A|\phi} \ket{v^0_A}, \\
    e^{-H_B t} \ket{\phi} \approx e^{-E^0_B t} \braket{v^0_B|\phi} \ket{v^0_B}.
\end{gather}
The problem arises from the fact that $\ket{v^0_A}$ and $\ket{v^0_B}$ appear with exponential factors of $e^{-E^0_A t}$ and $e^{-E^0_B t}$, respectively.  Calculations of the magnitude of the inner product $\braket{v^0_A|v^0_B}$ will suffer from large relative errors due to the factors of $e^{-E^0_A t}$ and $e^{-E^0_B t}$ for large $t$.

In Ref.~\cite{Sarkar:2023qjn}, an approach called the floating block method was introduced that addresses this problem.  The floating block method is based on the fact that
\begin{equation}
\lim_{t \rightarrow \infty}\frac{\braket{\phi|e^{-H_A t}e^{-H_B t}e^{-H_A t}e^{-H_B t}|\phi}}{\braket{\phi|e^{-2H_A t}e^{-2H_B t}|\phi}}=|\braket{v^0_A|v^0_B}|^2. \label{eq:FBequation}
\end{equation}
The problematic exponential factors of $e^{-E^0_A t}$ and $e^{-E^0_B t}$ cancel from this ratio.  We can compute the complex phase of the inner product using
\begin{equation}
\lim_{t \rightarrow \infty}\frac{\braket{\phi|e^{-2H_A t}e^{-2H_B t}|\phi}}{|\braket{\phi|e^{-2H_A t}e^{-2H_B t}|\phi}|}=\frac{\braket{v^0_A|v^0_B}}{|\braket{v^0_A|v^0_B}|}. \label{phase}
\end{equation}  
We use the phase convention that $\braket{v^0_A|\phi}$ and $\braket{v^0_B|\phi}$ are positive.

When performing Monte Carlo simulations for large systems, one prefers to compute ratios of quantities that are strongly correlated with each other.  Otherwise, statistical noise can become problematic, especially for large systems.  Let us define $Z(t_1,t_2,t_3,t_4)$ to be the amplitude
\begin{equation}
 Z(t_1,t_2,t_3,t_4) = \braket{\phi|e^{-H_A t_1}e^{-H_B t_2}e^{-H_A t_3}e^{-H_B t_4}|\phi}. \label{eq:t1t2t3t4}
\end{equation}
In the floating block method, we compute ratios
\begin{equation}
\frac{Z(t_1,t_2,t_3,t_4)}{Z(t'_1,t'_2,t'_3,t'_4)},
\end{equation}
for values $t_1,t_2,t_3,t_4$ and $t'_1,t'_2,t'_3,t'_4$ that are only slightly different from each other.  We compute the ratios,
\begin{equation}
\frac{Z(t_1,t_2,t_3,t_4)}{Z(t'_1,t'_2,t'_3,t'_4)}
\frac{Z(t'_1,t'_2,t'_3,t'_4)}{Z(t''_1,t''_2,t''_3,t''_4)}
\frac{Z(t''_1,t''_2,t''_3,t''_4)}{Z(t'''_1,t'''_2,t'''_3,t'''_4)} \cdots. \label{eq:ratios}
\end{equation}
We note that multiplying these ratios together forms a telescoping product where the adjacent numerator and denominator terms cancel.  Using this telescoping product, we can compute the ratio of the numerator and denominator in Eq.~\eqref{eq:FBequation}.  

The floating block method was used in Ref.~\cite{Sarkar:2023qjn} to compute the binding energies of ${}^{4}$He, ${}^{8}$Be, ${}^{12}$C, and ${}^{16}$O using auxiliary-field Monte Carlo lattice simulations.  The lattice Hamiltonian had the form $H_{\rm free} + c_LV_L + c_{NL}V_{NL}$, where $V_L$ is a two-nucleon interaction with local interactions and $V_{NL}$ is a two-nucleon interaction with nonlocal interactions.  This study confirmed the observation in Ref.~\cite{Elhatisari:2016owd} that symmetric nuclear matter sits close to a quantum phase transition.  When the range and strength of the local interactions are varied, symmetric nuclear matter can undergo a first-order transition between a nuclear liquid and a Bose gas of alpha particles. 

\section{Rank-one operator method for operator insertions}
In the auxiliary-field formalism, many-body amplitudes are determinants of matrices whose elements are composed of single-nucleon amplitudes.  When we insert a $k$-body operator $O$ in the midpoint of the Euclidean-time evolution in Eq.~\eqref{eq:expect}, this can be written as a sum of products of $k$ single-nucleon operators.  Consider a single term of the form $O = \prod_{m=1}^k o_m$, where each $o_m$ is a normal-ordered single-nucleon operator. We can insert the normal-ordered exponential $:\exp\left(\sum_{m=1}^k c_m o_m \right):$ in the midpoint of the Euclidean-time evolution and take one derivative with respect to each $c_m$ to obtain the desired product $O = \prod_{m=1}^k o_m$.  These derivatives can be computed using higher-order Jacobi formulas for derivatives of matrix determinants, or they can be evaluated with numerical differentiation using finite differences.  In either way, the scaling of the number of terms is exponential in $k$.  This is computationally expensive for larger values of $k$, and the problem is compounded if we also need to insert copies of the first-order transfer matrix $M'$ when using perturbation theory.

In Ref.~\cite{Ma:2023ahg}, a technique called the rank-one operator method was introduced to address this computational problem.  The rank of a matrix is defined as the number of linearly independent rows or columns.  We can decompose our $k$-body operator as a sum of terms $O = \prod_{m=1 \cdots k} o_m$, where each $o_m$ is a rank-one normal-ordered single-nucleon operator.  Now, when we insert the operator $:\exp\left(\sum_{m=1, \cdots k} c_m o_m \right):$ 
at the midpoint of the Euclidean-time evolution, the resulting many-body amplitude will have at most one power of $c_m$ for each $m$.  There are no higher powers of $c_m$ because $o_m$ is a rank-one operator and cannot annihilate or create more than one nucleon due to the Pauli exclusion principle.  We now simply take each $c_m$ to be numerically large and divide the many-body amplitude by $\prod_{m=1 \cdots k} c_m$.  This yields the desired amplitude for the insertion of $O = \prod_{m=1 \cdots k} o_m$.

The rank-one operator method was implemented in Ref.~\cite{Ma:2023ahg} to compute density and spin correlations in hot neutron gases using chiral EFT interactions at N3LO.  These correlations determine important properties of neutrino scattering and heating during core-collapse supernovae, and the NLEFT results can be used to improve supernova simulations.

\section{Summary and outlook}
Lattice EFT applies the principles of EFT in a lattice framework and has been used to describe the low-energy interactions of protons and neutrons.  In this review, we have covered several fundamental concepts, theoretical methods, and computational algorithms.  We have discussed basic tools such as the construction of the lattice Hamiltonian, lattice scattering with spherical wall boundaries, Euclidean-time projection, perturbation theory, and auxiliary-field Monte Carlo simulations.  We also reviewed more advanced techniques developed in recent years such as wavefunction matching, the pinhole algorithm, the pinhole trace algorithm, the adiabatic projection method, and the rank-one operator method.  Although we have focused on calculations of nuclei composed of protons and neutrons, there are also new developments in NLEFT related to simulations of hypernuclei containing one or more hyperons \cite{Frame:2020mvv,Hildenbrand:2022imw,Hildenbrand:2024ypw} and hyperneutron matter \cite{Tong:2024egi}.

As noted in the introduction, the methodology of lattice EFT is different from that of other {\it ab initio} approaches and can provide new information that is complementary to the other methods. 
 Several such examples have been illustrated in this review.  Every effort has been made to keep the material in this review as accessible and readable as possible.  Readers interested in learning more details about the various topics are encouraged to read the primary materials cited.    

\section*{DISCLOSURE STATEMENT}
The author is not aware of any affiliations, memberships, funding, or financial holdings that
might be perceived as affecting the objectivity of this review. 

\section*{ACKNOWLEDGMENTS}
The author is deeply grateful to have worked with many friends and colleagues who have contributed greatly to advances in lattice effective field theory.  They include: Jose Manuel Alarcón, Bu$\overline{\rm g}$ra Borasoy, Shahin Bour, Lukas Bovermann, Jiunn-Wei Chen, Dechuan Du, Evgeny Epelbaum, Serdar Elhatisari, Dillon Frame, Hans-Werner Hammer, Rongzheng He, Fabian Hildenbrand, Myungkuk Kim, Youngman Kim, Nico Klein, Sebastian König, Hermann Krebs, Ning Li, Bing-Nan Lu, Thomas Luu, Yuan-Zhuo Ma, Ulf-G. Mei{\ss}ner, Michelle Pine, Zhengxue Ren, Alexander Rokash, Gautam Rupak, Avik Sarkar, Thomas Schäfer, Shihang Shen, Young-Ho Song, Gianluca Stellin, Richard Thomson, Hui Tong, Thomas Vonk, Congwu Wang, Qian Wang, Teng Wang, Hang Yu, and Shuang Zhang.  Special thanks to Ulf-G. Mei{\ss}ner for providing comments on the manuscript. 
 Financial support is provided in part by the U.S. Department of Energy (grants DE-SC0013365, DE-SC0023175, DE-SC0023658, DE-SC0024586) and the U.S. National Science Foundation (grant PHY-2310620).  Computational resources provided by the Gauss Centre for Supercomputing e.V. (www.gauss-centre.eu) for computing time on the GCS Supercomputer JUWELS at J{\"u}lich Supercomputing Centre (JSC) and special GPU time allocated on JURECA-DC as well as
the Oak Ridge Leadership Computing Facility through the INCITE award ``Ab-initio nuclear structure and nuclear reactions.''

\bibliography{References}

\begin{thebibliography}{161}
\expandafter\ifx\csname natexlab\endcsname\relax\def\natexlab#1{#1}\fi

\bibitem{Navratil:2003ef}
Navratil P, Ormand WE.
\newblock \textit{Phys. Rev.} C68:034305 (2003)

\bibitem{Maris:2008ax}
Maris P, Vary JP, Shirokov AM.
\newblock \textit{Phys. Rev. C} 79:014308 (2009)

\bibitem{Barrett:2013nh}
Barrett BR, Navratil P, Vary JP.
\newblock \textit{Prog. Part. Nucl. Phys.} 69:131--181 (2013)

\bibitem{Wirth:2017bpw}
Wirth R, Gazda D, Navrátil P, Roth R.
\newblock \textit{Phys. Rev. C} 97(6):064315 (2018)

\bibitem{Navratil:2016ycn}
Navr\'atil P, Quaglioni S, Hupin G, Romero-Redondo C, Calci A.
\newblock \textit{Phys. Scripta} 91(5):053002 (2016)

\bibitem{Hupin:2018biv}
Hupin G, Quaglioni S, Navr\'atil P.
\newblock \textit{Nature Commun.} 10(1):351 (2019)

\bibitem{Hebborn:2022iiz}
Hebborn C, Hupin G, Kravvaris K, Quaglioni S, Navr\'atil P, Gysbers P.
\newblock \textit{Phys. Rev. Lett.} 129(4):042503 (2022)

\bibitem{Dreyfuss:2020lss}
Dreyfuss AC, Launey KD, Escher JE, Sargsyan GH, Baker RB, et~al.
\newblock \textit{Phys. Rev. C} 102(4):044608 (2020)

\bibitem{Becker:2023dqe}
Becker KS, Launey KD, Ekstr\"om A, Dytrych T.
\newblock \textit{Front. in Phys.} 11:1064601 (2023)

\bibitem{Carlson:2014vla}
Carlson J, Gandolfi S, Pederiva F, Pieper SC, Schiavilla R, et~al.
\newblock \textit{Rev. Mod. Phys.} 87:1067 (2015)

\bibitem{Dawkins:2019vcr}
Dawkins WG, Carlson J, van Kolck U, Gezerlis A.
\newblock \textit{Phys. Rev. Lett.} 124(14):143402 (2020)

\bibitem{King:2022zkz}
King GB, Baroni A, Cirigliano V, Gandolfi S, Hayen L, et~al.
\newblock \textit{Phys. Rev. C} 107(1):015503 (2023)

\bibitem{Lonardoni:2018nob}
Lonardoni D, Gandolfi S, Lynn J, Petrie C, Carlson J, et~al.
\newblock \textit{Phys. Rev. C} 97(4):044318 (2018)

\bibitem{Lovato:2022apd}
Lovato A, Bombaci I, Logoteta D, Piarulli M, Wiringa RB.
\newblock \textit{Phys. Rev. C} 105(5):055808 (2022)

\bibitem{Arthuis:2022ixv}
Arthuis P, Barbieri C, Pederiva F, Roggero A.
\newblock \textit{Phys. Rev. C} 107(4):044303 (2023)

\bibitem{Idini:2019hkq}
Idini A, Barbieri C, Navrátil P.
\newblock \textit{Phys. Rev. Lett.} 123(9):092501 (2019)

\bibitem{Carbone:2019pkr}
Carbone A, Schwenk A.
\newblock \textit{Phys. Rev. C} 100(2):025805 (2019)

\bibitem{Barbieri:2021ezv}
Barbieri C, Duguet T, Som\`a V.
\newblock \textit{Phys. Rev. C} 105(4):044330 (2022)

\bibitem{Tichai:2020dna}
Tichai A, Roth R, Duguet T.
\newblock \textit{Front. in Phys.} 8:164 (2020)

\bibitem{Demol:2020mzd}
Demol P, Frosini M, Tichai A, Som\`a V, Duguet T.
\newblock \textit{Annals Phys.} 424:168358 (2021)

\bibitem{Frosini:2021ddm}
Frosini M, Duguet T, Ebran JP, Bally B, Hergert H, et~al.
\newblock \textit{Eur. Phys. J. A} 58(4):64 (2022)

\bibitem{Holt:2019gmc}
Stroberg SR, Holt JD, Schwenk A, Simonis J.
\newblock \textit{Phys. Rev. Lett.} 126(2):022501 (2021)

\bibitem{Stroberg:2019mxo}
Stroberg SR, Bogner SK, Hergert H, Holt JD.
\newblock \textit{Ann. Rev. Nucl. Part. Sci.} 69:307--362 (2019)

\bibitem{Yao:2019rck}
Yao J, Bally B, Engel J, Wirth R, Rodríguez T, Hergert H.
\newblock \textit{Phys. Rev. Lett.} 124(23):232501 (2020)

\bibitem{Hu:2021trw}
Hu B, et~al.
\newblock \textit{Nature Phys.} 18(10):1196--1200 (2022)

\bibitem{Sun:2018fmu}
Sun Z, Morris T, Hagen G, Jansen G, Papenbrock T.
\newblock \textit{Phys. Rev. C} 98(5):054320 (2018)

\bibitem{Jiang:2020the}
Jiang WG, Ekstr\"om A, Forss\'en C, Hagen G, Jansen GR, Papenbrock T.
\newblock \textit{Phys. Rev. C} 102(5):054301 (2020)

\bibitem{Frosini:2021fjf}
Frosini M, Duguet T, Ebran JP, Som\`a V.
\newblock \textit{Eur. Phys. J. A} 58(4):62 (2022)

\bibitem{Frosini:2021sxj}
Frosini M, Duguet T, Ebran JP, Bally B, Mongelli T, et~al.
\newblock \textit{Eur. Phys. J. A} 58(4):63 (2022)

\bibitem{Otsuka:2022bcf}
Otsuka T, Abe T, Yoshida T, Tsunoda Y, Shimizu N, et~al.
\newblock \textit{Nature Commun.} 13(1):2234 (2022)

\bibitem{Weinberg:1990rz}
Weinberg S.
\newblock \textit{Phys. Lett.} B251:288--292 (1990)

\bibitem{Weinberg:1991um}
Weinberg S.
\newblock \textit{Nucl. Phys.} B363:3--18 (1991)

\bibitem{Weinberg:1992yk}
Weinberg S.
\newblock \textit{Phys. Lett.} B295:114--121 (1992)

\bibitem{Ordonez:1992xp}
Ordonez C, van Kolck U.
\newblock \textit{Phys. Lett.} B291:459--464 (1992)

\bibitem{Ordonez:1993tn}
Ordonez C, Ray L, van Kolck U.
\newblock \textit{Phys. Rev. Lett.} 72:1982--1985 (1994)

\bibitem{Friar:1994}
Friar JL, Coon SA.
\newblock \textit{Phys. Rev.} C49:1272--1280 (1994)

\bibitem{vanKolck:1994yi}
van Kolck U.
\newblock \textit{Phys. Rev.} C49:2932--2941 (1994)

\bibitem{Bernard:1995dp}
Bernard V, Kaiser N, Mei{\ss}ner UG.
\newblock \textit{Int. J. Mod. Phys.} E4:193--346 (1995)

\bibitem{Ordonez:1996rz}
Ordonez C, Ray L, van Kolck U.
\newblock \textit{Phys. Rev.} C53:2086--2105 (1996)

\bibitem{Kaiser:1997mw}
Kaiser N, Brockmann R, Weise W.
\newblock \textit{Nucl. Phys.} A625:758--788 (1997)

\bibitem{Friar:1998zt}
Friar JL, Huber D, van Kolck U.
\newblock \textit{Phys. Rev.} C59:53--58 (1999)

\bibitem{Epelbaum:1998na}
Epelbaum E, Glockle W, Kruger A, Mei{\ss}ner UG.
\newblock \textit{Nucl. Phys.} A645:413--438 (1999)

\bibitem{Epelbaum:1998hg}
Epelbaum E, Glockle W, Mei{\ss}ner UG.
\newblock \textit{Phys. Lett.} B439:1--5 (1998)

\bibitem{Epelbaum:1998ka}
Epelbaum E, Glockle W, Mei{\ss}ner UG.
\newblock \textit{Nucl. Phys.} A637:107--134 (1998)

\bibitem{Epelbaum:2008ga}
Epelbaum E, Hammer HW, Mei{\ss}ner UG.
\newblock \textit{Rev. Mod. Phys.} 81:1773 (2009)

\bibitem{Machleidt:2016rvv}
Machleidt R, Sammarruca F.
\newblock \textit{Phys. Scripta} 91(8):083007 (2016)

\bibitem{Hammer:2019poc}
Hammer HW, K\"onig S, van Kolck U.
\newblock \textit{Rev. Mod. Phys.} 92(2):025004 (2020)

\bibitem{Muller:1999cp}
M{\"u}ller HM, Koonin SE, Seki R, van Kolck U.
\newblock \textit{Phys. Rev.} C61:044320 (2000)

\bibitem{Shushpanov:1998ms}
Shushpanov IA, Smilga AV.
\newblock \textit{Phys. Rev.} D59:054013 (1999)

\bibitem{Lewis:2000cc}
Lewis R, Ouimet PPA.
\newblock \textit{Phys. Rev.} D64:034005 (2001)

\bibitem{Chandrasekharan:2003wy}
Chandrasekharan S, Pepe M, Steffen FD, Wiese UJ.
\newblock \textit{JHEP} 12:035 (2003)

\bibitem{Lee:2004si}
Lee D, Borasoy B, Sch{\"a}fer T.
\newblock \textit{Phys. Rev.} C70:014007 (2004)

\bibitem{Lee:2004qd}
Lee D, Sch{\"a}fer T.
\newblock \textit{Phys. Rev.} C72:024006 (2005)

\bibitem{Lee:2005is}
Lee D, Sch{\"a}fer T.
\newblock \textit{Phys. Rev.} C73:015201 (2006)

\bibitem{Lee:2005it}
Lee D, Sch{\"a}fer T.
\newblock \textit{Phys. Rev.} C73:015202 (2006)

\bibitem{Borasoy:2005yc}
Borasoy B, Krebs H, Lee D, Mei{\ss}ner UG.
\newblock \textit{Nucl. Phys.} A768:179--193 (2006)

\bibitem{Chen:2003vy}
Chen JW, Kaplan DB.
\newblock \textit{Phys. Rev. Lett.} 92:257002 (2004)

\bibitem{Bulgac:2005pj}
Bulgac A, Drut JE, Magierski P.
\newblock \textit{Phys. Rev. Lett.} 96:090404 (2006)

\bibitem{Burovski:2006}
{Burovski} E, {Prokof'ev} N, {Svistunov} B, {Troyer} M.
\newblock \textit{New Journal of Physics} 8(8):153 (2006)

\bibitem{Wingate:2006wy}
Wingate M.
\newblock \textit{PoS} LAT2006:153 (2006)

\bibitem{Abe:2007fe}
Abe T, Seki R.
\newblock \textit{Phys. Rev.} C79:054002 (2009)

\bibitem{Alexandru:2019gmp}
Alexandru A, Bedaque PF, Warrington NC.
\newblock \textit{Phys. Rev. C} 101(4):045805 (2020)

\bibitem{Alexandru:2020zti}
Alexandru A, Bedaque P, Berkowitz E, Warrington NC.
\newblock \textit{Phys. Rev. Lett.} 126(13):132701 (2021)

\bibitem{Borasoy:2006qn}
Borasoy B, Epelbaum E, Krebs H, Lee D, Mei{\ss}ner UG.
\newblock \textit{Eur. Phys. J.} A31:105--123 (2007)

\bibitem{Lee:2008fa}
Lee D.
\newblock \textit{Prog. Part. Nucl. Phys.} 63:117--154 (2009)

\bibitem{Drut:2012md}
Drut JE, Nicholson AN.
\newblock \textit{J. Phys. G} 40:043101 (2013)

\bibitem{Lee:2016fhn}
Lee D.
\newblock \textit{Lect. Notes Phys.} 936:237--261 (2017)

\bibitem{Lahde:2019npb}
Lähde TA, Meißner UG.
\newblock \textit{Lect. Notes Phys.} 957:1--396 (2019)

\bibitem{Korber:2015rce}
K\"orber C, Luu T.
\newblock \textit{Phys. Rev. C} 93(5):054002 (2016)

\bibitem{Lu:2019nbg}
Lu BN, Li N, Elhatisari S, Lee D, Drut JE, et~al.
\newblock \textit{Phys. Rev. Lett.} 125(19):192502 (2020)

\bibitem{Song:2021yst}
Song YH, Kim Y, Li N, Lu BN, He R, Lee D.
\newblock \textit{Phys. Rev. C} 104(4):044304 (2021)

\bibitem{Bulgac:2005a}
Bulgac A, Drut JE, Magierski P.
\newblock \textit{Phys. Rev. Lett.} 96:090404 (2006)

\bibitem{Li:2018ymw}
Li N, Elhatisari S, Epelbaum E, Lee D, Lu BN, Meißner UG.
\newblock \textit{Phys. Rev.} C98(4):044002 (2018)

\bibitem{Epelbaum:2009rkz}
Epelbaum E, Krebs H, Lee D, Mei{\ss}ner UG.
\newblock \textit{Eur. Phys. J. A} 40:199--213 (2009)

\bibitem{Epelbaum:2011md}
Epelbaum E, Krebs H, Lee D, Mei{\ss}ner UG.
\newblock \textit{Phys. Rev. Lett.} 106:192501 (2011)

\bibitem{Epelbaum:2012qn}
Epelbaum E, Krebs H, L{\"a}hde T, Lee D, Mei{\ss}ner UG.
\newblock \textit{Phys. Rev. Lett.} 109:252501 (2012)

\bibitem{Epelbaum:2013paa}
Epelbaum E, Krebs H, L{\"a}hde TA, Lee D, Mei{\ss}ner UG, Rupak G.
\newblock \textit{Phys. Rev. Lett.} 112(10):102501 (2014)

\bibitem{Alarcon:2017zcv}
Alarc\'on JM, Du D, Klein N, L\"ahde TA, Lee D, et~al.
\newblock \textit{Eur. Phys. J. A} 53(5):83 (2017)

\bibitem{Li:2019ldq}
Li N, Elhatisari S, Epelbaum E, Lee D, Lu B, Meißner UG.
\newblock \textit{Phys. Rev.} C99(6):064001 (2019)

\bibitem{Luscher:1986pf}
L{\"u}scher M.
\newblock \textit{Commun. Math. Phys.} 105:153--188 (1986)

\bibitem{Luscher:1990ck}
L{\"u}scher M, Wolff U.
\newblock \textit{Nucl. Phys.} B339:222--252 (1990)

\bibitem{Luscher:1990ux}
L{\"u}scher M.
\newblock \textit{Nucl. Phys.} B354:531--578 (1991)

\bibitem{Carlson:1984zz}
Carlson J, Pandharipande V, Wiringa R.
\newblock \textit{Nucl. Phys. A} 424(1):47 -- 59 (1984)

\bibitem{Borasoy:2007vy}
Borasoy B, Epelbaum E, Krebs H, Lee D, Mei{\ss}ner UG.
\newblock \textit{Eur. Phys. J.} A34:185--196 (2007)

\bibitem{Lu:2015riz}
Lu BN, L{\"a}hde TA, Lee D, Mei{/ss}ner UG.
\newblock \textit{Phys. Lett.} B760:309--313 (2016)

\bibitem{Bovermann:2019jbt}
Bovermann L, Epelbaum E, Krebs H, Lee D.
\newblock \textit{Phys. Rev. C} 100(6):064001 (2019)

\bibitem{Epelbaum:2014efa}
Epelbaum E, Krebs H, Mei\ss{}ner UG.
\newblock \textit{Eur. Phys. J. A} 51(5):53 (2015)

\bibitem{Epelbaum:2014sza}
Epelbaum E, Krebs H, Mei\ss{}ner UG.
\newblock \textit{Phys. Rev. Lett.} 115(12):122301 (2015)

\bibitem{Furnstahl:2014xsa}
Furnstahl RJ, Phillips DR, Wesolowski S.
\newblock \textit{J. Phys. G} 42(3):034028 (2015)

\bibitem{Furnstahl:2015rha}
Furnstahl RJ, Klco N, Phillips DR, Wesolowski S.
\newblock \textit{Phys. Rev. C} 92(2):024005 (2015)

\bibitem{Stoks:1993tb}
Stoks VGJ, Kompl RAM, Rentmeester MCM, de~Swart JJ.
\newblock \textit{Phys. Rev.} C48:792--815 (1993)

\bibitem{Lu:2021tab}
Lu BN, Li N, Elhatisari S, Ma YZ, Lee D, Mei\ss{}ner UG.
\newblock \textit{Phys. Rev. Lett.} 128(24):242501 (2022)

\bibitem{Curry:2024gcz}
Curry R, Somasundaram R, Gandolfi S, Gezerlis A, Tews I.
\newblock \textit{Phys. Rev. C} 111(1):015801 (2025)

\bibitem{Hubbard:1959ub}
Hubbard J.
\newblock \textit{Phys. Rev. Lett.} 3:77--80 (1959)

\bibitem{Stratonovich:1958}
Stratonovich RL.
\newblock \textit{Soviet Phys. Doklady} 2:416--419 (1958)

\bibitem{Koonin:1986}
Koonin SE.
\newblock \textit{Journal of Statistical Physics} 43(5-6):985--990 (1986)

\bibitem{Chen:2004rq}
Chen JW, Lee D, Sch{\"a}fer T.
\newblock \textit{Phys. Rev. Lett.} 93:242302 (2004)

\bibitem{Korber:2017emn}
K\"orber C, Berkowitz E, Luu T.
\newblock \textit{EPL} 119(6):60006 (2017)

\bibitem{Lu:2018bat}
Lu BN, Li N, Elhatisari S, Lee D, Epelbaum E, Meißner UG.
\newblock \textit{Phys. Lett.} B797:134863 (2019)

\bibitem{Wigner:1936dx}
Wigner E.
\newblock \textit{Phys. Rev.} 51:106--119 (1937)

\bibitem{Kaplan:1995yg}
Kaplan DB, Savage MJ.
\newblock \textit{Phys. Lett.} B365:244--251 (1996)

\bibitem{Kaplan:1996rk}
Kaplan DB, Manohar AV.
\newblock \textit{Phys. Rev.} C56:76--83 (1997)

\bibitem{Mehen:1999qs}
Mehen T, Stewart IW, Wise MB.
\newblock \textit{Phys. Rev. Lett.} 83:931--934 (1999)

\bibitem{Banerjee:2001js}
Banerjee MK, Cohen TD, Gelman BA.
\newblock \textit{Phys. Rev. C} 65:034011 (2002)

\bibitem{CalleCordon:2008cz}
Calle~Cordon A, Ruiz~Arriola E.
\newblock \textit{Phys. Rev.} C78:054002 (2008)

\bibitem{Lee:2020esp}
Lee D, et~al.
\newblock \textit{Phys. Rev. Lett.} 127(6):062501 (2021)

\bibitem{Lee:2007eu}
Lee D.
\newblock \textit{Phys. Rev. Lett.} 98:182501 (2007)

\bibitem{Lee:2004ze}
Lee D.
\newblock \textit{Phys. Rev.} C70:064002 (2004)

\bibitem{Lee:2004hc}
Lee D.
\newblock \textit{Phys. Rev.} C71:044001 (2005)

\bibitem{Hoffman:2016jqv}
Hoffman MD, Loheac AC, Porter WJ, Drut JE.
\newblock \textit{Phys. Rev. A} 95(3):033602 (2017)

\bibitem{Lahde:2015ona}
L{\"a}hde TA, Luu T, Lee D, Mei{\ss}ner UG, Epelbaum E, et~al.
\newblock \textit{Eur. Phys. J.} A51(7):92 (2015)

\bibitem{Wlazlowski:2014jna}
Wlaz{\l}owski G, Holt JW, Moroz S, Bulgac A, Roche KJ.
\newblock \textit{Phys. Rev. Lett.} 113(18):182503 (2014)

\bibitem{Frame:2017fah}
Frame D, He R, Ipsen I, Lee D, Lee D, Rrapaj E.
\newblock \textit{Phys. Rev. Lett.} 121(3):032501 (2018)

\bibitem{Borasoy:2007vk}
Borasoy B, Epelbaum E, Krebs H, Lee D, Mei{\ss}ner UG.
\newblock \textit{Eur. Phys. J.} A35:357--367 (2008)

\bibitem{Epelbaum:2009zsa}
Epelbaum E, Krebs H, Lee D, Mei{\ss}ner UG.
\newblock \textit{Eur. Phys. J. A} 41:125--139 (2009)

\bibitem{Epelbaum:2009pd}
Epelbaum E, Krebs H, Lee D, Mei{\ss}ner UG.
\newblock \textit{Phys. Rev. Lett.} 104:142501 (2010)

\bibitem{Epelbaum:2010xt}
Epelbaum E, Krebs H, Lee D, Mei{\ss}ner UG.
\newblock \textit{Eur. Phys. J.} A45:335--352 (2010)

\bibitem{Lahde:2013uqa}
L{\"a}hde TA, Epelbaum E, Krebs H, Lee D, Mei{\ss}ner UG, Rupak G.
\newblock \textit{Phys. Lett.} B732:110--115 (2014)

\bibitem{Shen:2022bak}
Shen S, Elhatisari S, L\"ahde TA, Lee D, Lu BN, Mei\ss{}ner UG.
\newblock \textit{Nature Commun.} 14(1):2777 (2023)

\bibitem{Gnech:2023prs}
Gnech A, Fore B, Tropiano AJ, Lovato A.
\newblock \textit{Phys. Rev. Lett.} 133(14):142501 (2024)

\bibitem{Shen:2022arg}
Shen S, L\"ahde TA, Lee D, Mei\ss{}ner UG.
\newblock \textit{PoS} LATTICE2022:241 (2023)

\bibitem{Shen:2024qzi}
Shen S, Elhatisari S, Lee D, Mei\ss{}ner UG, Ren Z.
\newblock \textit{arXiv:2411.14935}  (2024)

\bibitem{Meissner:2023cvo}
Mei\ss{}ner UG, Shen S, Elhatisari S, Lee D.
\newblock \textit{Phys. Rev. Lett.} 132(6):062501 (2024)

\bibitem{Elhatisari:2022zrb}
Elhatisari S, et~al.
\newblock \textit{Nature} 630(8015):59--63 (2024)

\bibitem{Ma:2023ahg}
Ma YZ, Lin Z, Lu BN, Elhatisari S, Lee D, et~al.
\newblock \textit{Phys. Rev. Lett.} 132(23):232502 (2024)

\bibitem{Konig:2023rwe}
K\"onig K, et~al.
\newblock \textit{Phys. Rev. Lett.} 132(16):162502 (2024), [Erratum:
  Phys.Rev.Lett. 133, 059901 (2024)]

\bibitem{Zhang:2024wfd}
Zhang S, Elhatisari S, Mei\ss{}ner UG, Shen S.
\newblock \textit{arXiv:2411.17462}  (2024)

\bibitem{Elhatisari:2024otn}
Elhatisari S, Hildenbrand F, Mei\ss{}ner UG.
\newblock \textit{Phys. Lett. B} 859:139086 (2024)

\bibitem{Elhatisari:2017eno}
Elhatisari S, Epelbaum E, Krebs H, Lähde TA, Lee D, et~al.
\newblock \textit{Phys. Rev. Lett.} 119(22):222505 (2017)

\bibitem{Shen:2021kqr}
Shen S, L\"ahde TA, Lee D, Mei\ss{}ner UG.
\newblock \textit{Eur. Phys. J. A} 57(9):276 (2021)

\bibitem{Summerfield:2021oex}
Summerfield N, Lu BN, Plumberg C, Lee D, Noronha-Hostler J, Timmins A.
\newblock \textit{Phys. Rev. C} 104(4):L041901 (2021)

\bibitem{Ding:2023ibq}
Ding C, Pang LG, Zhang S, Ma YG.
\newblock \textit{Chin. Phys. C} 47(2):024105 (2023)

\bibitem{Zhao:2024feh}
Zhao XL, Ma GL, Zhou Y, Lin ZW, Zhang C.
\newblock \textit{arXiv:}  (2024)

\bibitem{Zhang:2024vkh}
Zhang C, Chen J, Giacalone G, Huang S, Jia J, Ma YG.
\newblock \textit{arXiv:2404.08385}  (2024)

\bibitem{Giacalone:2024luz}
Giacalone G, et~al.
\newblock \textit{arXiv:2402.05995}  (2024)

\bibitem{Giacalone:2024ixe}
Giacalone G, et~al.
\newblock \textit{arXiv:2405.20210}  (2024)

\bibitem{Elliott:2013pna}
Elliott JB, Lake PT, Moretto LG, Phair L.
\newblock \textit{Phys. Rev. C} 87(5):054622 (2013)

\bibitem{Ren:2023ued}
Ren Z, Elhatisari S, L\"ahde TA, Lee D, Mei\ss{}ner UG.
\newblock \textit{Phys. Lett. B} 850:138463 (2024)

\bibitem{Pine:2013zja}
Pine M, Lee D, Rupak G.
\newblock \textit{Eur. Phys. J.} A49:151 (2013)

\bibitem{Rokash:2015hra}
Rokash A, Pine M, Elhatisari S, Lee D, Epelbaum E, Krebs H.
\newblock \textit{Phys. Rev.} C92(5):054612 (2015)

\bibitem{Elhatisari:2015iga}
Elhatisari S, Lee D, Rupak G, Epelbaum E, Krebs H, et~al.
\newblock \textit{Nature} 528:111 (2015)

\bibitem{Elhatisari:2016hby}
Elhatisari S, Lee D, Mei{\ss}ner UG, Rupak G.
\newblock \textit{Eur. Phys. J.} A52(6):174 (2016)

\bibitem{Elhatisari:2019fvk}
Elhatisari S.
\newblock \textit{Eur. Phys. J. A} 55(8):144 (2019)

\bibitem{Elhatisari:2021eyg}
Elhatisari S, L\"ahde TA, Lee D, Mei\ss{}ner UG, Vonk T.
\newblock \textit{JHEP} 02:001 (2022)

\bibitem{Rupak:2013aue}
Rupak G, Lee D.
\newblock \textit{Phys. Rev. Lett.} 111(3):032502 (2013)

\bibitem{Frame:2019jsw}
Frame DK. 2019.
\newblock {Ab Initio Simulations of Light Nuclear Systems Using Eigenvector
  Continuation and Auxiliary Field Monte Carlo}.
\newblock Doctoral thesis, Michigan State University

\bibitem{Demol:2019yjt}
Demol P, Duguet T, Ekstr\"om A, Frosini M, Hebeler K, et~al.
\newblock \textit{Phys. Rev. C} 101(4):041302 (2020)

\bibitem{Konig:2019adq}
K\"onig S, Ekstr\"om A, Hebeler K, Lee D, Schwenk A.
\newblock \textit{Phys. Lett. B} 810:135814 (2020)

\bibitem{Ekstrom:2019lss}
Ekstr\"om A, Hagen G.
\newblock \textit{Phys. Rev. Lett.} 123(25):252501 (2019)

\bibitem{Furnstahl:2020abp}
Furnstahl RJ, Garcia AJ, Millican PJ, Zhang X.
\newblock \textit{Phys. Lett. B} 809:135719 (2020)

\bibitem{Bonilla:2022rph}
Bonilla E, Giuliani P, Godbey K, Lee D.
\newblock \textit{Phys. Rev. C} 106(5):054322 (2022)

\bibitem{Melendez:2022kid}
Melendez JA, Drischler C, Furnstahl RJ, Garcia AJ, Zhang X.
\newblock \textit{J. Phys. G} 49(10):102001 (2022)

\bibitem{hesthaven2015certified}
Hesthaven J, Rozza G, Stamm B.
\newblock SpringerBriefs in Mathematics. Springer International Publishing
  (2016)

\bibitem{Quarteroni:218966}
Quarteroni A, Manzoni A, Negri F.
\newblock La Matematica per il 3+2. 92. Springer International Publishing
  (2016)

\bibitem{Duguet:2023wuh}
Duguet T, Ekstr\"om A, Furnstahl RJ, K\"onig S, Lee D.
\newblock \textit{Rev. Mod. Phys.} 96(3):031002 (2024)

\bibitem{Sarkar:2023qjn}
Sarkar A, Lee D, Mei\ss{}ner UG.
\newblock \textit{Phys. Rev. Lett.} 131(24):242503 (2023)

\bibitem{Elhatisari:2016owd}
Elhatisari S, et~al.
\newblock \textit{Phys. Rev. Lett.} 117(13):132501 (2016)

\bibitem{Frame:2020mvv}
Frame D, L\"ahde TA, Lee D, Mei{\ss}ner UG.
\newblock \textit{Eur. Phys. J. A} 56(10):248 (2020)

\bibitem{Hildenbrand:2022imw}
Hildenbrand F, Elhatisari S, L\"ahde TA, Lee D, Mei\ss{}ner UG.
\newblock \textit{Eur. Phys. J. A} 58(9):167 (2022)

\bibitem{Hildenbrand:2024ypw}
Hildenbrand F, Elhatisari S, Ren Z, Mei\ss{}ner UG.
\newblock \textit{Eur. Phys. J. A} 60(10):215 (2024)

\bibitem{Tong:2024egi}
Tong H, Elhatisari S, Mei\ss{}ner UG.
\newblock \textit{arXiv:2405.01887}  (2024)

\end{thebibliography}
\bibliographystyle{ar-style5.bst}

\end{document}